\documentclass[useAMS,usenatbib]{mn2e}
\usepackage{lscape}
\usepackage{amsmath}
\usepackage{latexsym}
\usepackage{graphicx}
\usepackage{amssymb}

\newcommand{\lya}{Ly$\alpha$}

\newcommand{\zzz}{$z\sim3$}

\newcommand{\kpc}{$h^{-1}_{70}$ kpc}
\newcommand{\Mpc}{$h^{-1}_{70}$ Mpc}
\title[LBG ultraviolet features and environment] {Nurturing Lyman
Break Galaxies: Observed link between environment and spectroscopic
features}

\author[Cooke, Omori, \& Ryan-Weber]{J. Cooke$^1$\thanks{E-mail:
jcooke@astro.swin.edu.au}, Y. Omori$^2$, \& E. Ryan-Weber$^1$\\  
$^1$Centre for Astrophysics and Supercomputing, Swinburne University
of Technology, Hawthorn, VIC, 3122, Australia\\ 
$^2$Department of Physics, McGill University, Montreal, QC H2A 2T8,
Canada}

\begin{document}

\date{Accepted 0000 January 00. Received 0000 January 00; in original
form 2011 April 10}
\pagerange{\pageref{firstpage}--\pageref{lastpage}} \pubyear{0000}

\maketitle

\label{firstpage}
 
\begin{abstract} We examine the effects of magnitude, colour, and
\lya\ equivalent width (EW) on the spatial distribution of \zzz\ Lyman
break galaxies (LBGs) and report significant differences in the
two-point auto-correlation functions.  The results are obtained using
samples of $\sim$10,000--57,000 LBGs from the Canada-France-Hawaii
Telescope Legacy Survey Deep fields.  We find that magnitude has a
larger effect on the auto-correlation function amplitude on small
scales ($\lesssim$1 \Mpc, the one-halo term) and that colour is more
influential on large scales ($\gtrsim$1 \Mpc, the two-halo term).  We
find the most significant differences between auto-correlation
functions for LBGs with dominant net \lya\ EW in absorption (aLBGs)
and dominant net \lya\ EW in emission (eLBGs) determined from
$\gtrsim$95\% pure samples of each population using a photometric
technique calibrated from $\sim$1000 spectra.  The aLBG
auto-correlation function has a higher two-halo amplitude than the
full LBG sample and has a one-halo term departure from a power law fit
near $\sim$1 \Mpc, corresponding to the virial radii of
$M_{DM}\sim$10$^{13} M_\odot$ dark matter haloes.  In contrast, the
eLBG auto-correlation function has a one-halo term departure at
$\sim$0.12 \Mpc, suggesting parent haloes of $M_{DM}\sim$10$^{11}
M_\odot$, and a two-halo term that exhibits a curious ``hump'' on
intermediate scales that we localize to the faintest, bluest members.
The aLBG-eLBG cross-correlation function exhibits an anti-correlation
component that reinforces different physical locations for a
significant fraction of aLBGs and eLBGs.  We introduce a ``shell''
model for the eLBG auto-correlation function and find that the form
can be reproduced assuming a significant fraction of eLBGs have a
shell-like spatial distribution.  Based on the analysis of all LBG
sub-samples, and considering the natural asymmetric distribution of
LBGs on the colour-magnitude diagram, we conclude that aLBGs are more
likely to reside in group-like environments hosting multiple luminous
($i'<$ 26.4) LBGs whereas eLBGs are more likely to be found on group
outskirts and in the field.  Because \lya\ is a tracer of several
intrinsic properties, including morphology, the results presented here
imply that the mechanisms behind the morphology-density relation at
low redshift are in place at \zzz\ and that \lya\ EW may be a key
environment diagnostic.  Finally, our results show that the LBG
auto-correlation function amplitude is lower than the true average as
a result of the spatial anti-correlation of the spectral types.  This
results holds broad consequences for all auto-correlation functions
measured for any population that contains members residing in
different environments as the average amplitude, and hence the
inferred average dark matter mass, will always be underestimated.

\end{abstract}


\section{INTRODUCTION}\label{intro}

Lyman break galaxies (LBGs) are star forming galaxies at high redshift
detected by their strong ultraviolet (UV) continua and drop in flux
blueward of the Lyman limit \citep{steidel96}.  Although searches for
galaxies using other selection criteria and wavelengths have been
successful in finding various populations
\citep[e.g.,][]{vandokkum03,daddi04,chapman05}, LBGs are considered to
comprise the bulk of star forming galaxies at high redshift
\citep{reddy05,marchesini07}.

The spatial clustering of LBGs reveal that they reside in overdense
regions of the universe \citep[e.g.,][]
{steidel98,foucaud03,adelberger05,cooke06,hildebrandt09,bielby11}.
The clustering is typically quantified by the two-point correlation
function which is observed to closely follow a power law at large
scales (greater than $\sim$200 \kpc, physical), the so-called
`two-halo' term, that probe the separations of parent dark matter
haloes.  Surveys that probe the correlation function down to small
scales \citep[$\lesssim$50 \kpc, physical;][]{ouchi05,lee06}, the
so-called `one-halo' term, find a departure from a power law that
provides insight into the distribution of luminous galaxies (or
luminous sub-haloes) within parent dark matter haloes.

Relationships between the spatial distribution and magnitude of LBGs
has been reported in previous surveys and indicate that more luminous
LBGs are more strongly clustered
\citep{giavalisco01,ouchi04,kashikawa06}.  Here, we explore the
spatial distribution of LBGs divided into independent subsets based on
their magnitude, colour, and spectroscopic features and measure the
two-point auto-correlation and cross-correlation functions across both
the one-halo and two-halo scales.

Investigation into a relationship between clustering and spectroscopic
features is motivated by (1) the trend in magnitude with \lya\
equivalent width (EW) and the relationships between \lya\ EW and other
ultraviolet spectroscopic and morphological properties and (2) the
observed relationship between \lya\ EW and LBG pair separation.
\citet{shapley03} examine the spectra of $\sim$800 \zzz\ LBGs and find
an average luminosity increase with decreasing \lya\ EW.  In addition,
that work uncovers strong relationships between \lya\ EW and other
properties such as UV continua slope, star formation rate, low- and
high-ionization ISM absorption line EWs, and line velocity offsets
with respect to systemic redshifts (a potential outflow signature).
In addition, \cite{cooke09a} investigate the behaviour of \lya\ EW on
the colour-magnitude diagram and find a separation and asymmetric
distribution of \lya\ EW with colour and magnitude.  Red LBGs
typically exhibit dominant \lya\ in absorption and blue LBGs typically
show dominant \lya\ in emission.  The bulk of luminous LBGs are redder
systems exhibiting dominant \lya\ in absorption, i.e., there are few
luminous blue LBGs and fewer bright LBGs with dominant \lya\ in
emission.  Faint LBGs may consist of LBGs of both types, however,
spectroscopically confirmed faint LBGs are dominated by blue systems
that display \lya\ in emission.

The spectroscopic and spectroscopic/photometric close pairs studied in
\citet{cooke10} reveal that \zzz\ LBGs within $\lesssim20$ \kpc,
physical, of another LBG exhibit dominant \lya\ in emission.  This
fraction decreases with increasing separation and drops to
$\sim$50--60\% at $\gtrsim$50 \kpc, equivalent to the fraction
measured for the full \zzz\ population.  In addition, that work
introduced trends in morphology with \lya\ EW as interpreted from
Hubble Space Telescope (HST) restframe UV images and the
non-parametric analysis of \citet{law07}.  Specifically, LBGs with
dominant \lya\ in absorption are more often diffuse, extended (lower
Gini coefficients), and typically exhibit multiple star forming clumps
whereas LBGs with dominant \lya\ in emission are typically compact
(higher Gini coefficients) with an apparent single, typically strong,
star forming component (or two).  These trends are reinforced by
results of \citet{law11} that analyse LBGs in HST restframe optical
images where morphology is better understood.  Consequently, an
exploration of the large and small-scale correlation functions of LBGs
based on \lya\ EW, and thus their UV spectral properties and
morphology, provides a powerful means to investigate the interplay
between environment and galaxy properties at high redshift.

The auto-correlation function measures the clustering strength for a
galaxy population which can provide the bias of luminous galaxies with
respect to the underlying dark matter to infer average halo masses
and, when modeled with halo occupation distribution models, can
provide an estimate of the average number of luminous galaxies hosted
by the parent haloes.  In contrast, the cross-correlation function is
sensitive to differences in the spatial distributions of two
populations indicating whether or not the compared populations reside
in the same physical regions of the Universe.  In order to measure the
correlation functions over the range of separations necessary to
sample both the one- and two-halo regimes (a few kpc to tens of Mpc)
in a statistically meaningful way requires large ($\gtrsim10^4$),
wide-field samples.  Thus, examining LBGs based on their spectroscopic
properties requires an equivalent number of deep spectra which is
difficult to obtain using existing facilities.  Instead, we apply the
\zzz\ LBG spectral-type selection approach of \citet[][hereafter
C09]{cooke09a} to the four square-degree Deep fields of the
Canada-France-Hawaii Telescope Legacy Survey (CFHTLS) images, that
enables us to achieve the necessary large samples.  The photometric
spectral-type criteria are found to cleanly isolate two LBG subsets,
one with dominant \lya\ in absorption and the other with dominant
\lya\ in emission and their respective UV spectral properties with
$\gtrsim$95\% purity as determined from $\sim$1000 \zzz\ spectra.
Here we use $\sim$70 Keck spectra of the \zzz\ LBGs used here as a
confirmation of the criteria selection and purity (\S\ref{obs}).

The magnitude, colour, and spectral type auto-correlation functions
presented here unearth fundamental differences in their behaviour,
with the largest effect seen for the spectral types.  The spectral
type cross-correlation function exhibits an anti-correlation component
which indicates that a significant fraction of the two populations do
not reside in similar physical locations.  The results and tests
presented here point to a strong connection between the observed
internal properties of LBGs and external group and field environments.
Our analysis helps to provide order to the complex UV morphology of
LBGs and may provide links between LBG spectral properties,
environment, and kinematics to be investigated in a forthcoming paper.

This paper is organised as follows.  We discuss the observations in
$\S\ref{obs}$ and define our \zzz\ LBG galaxy selection and LBG
sub-samples in $\S\ref{z3colors}$.  Correlation functions and tests
are presented in $\S\ref{results}$ and are analysed over the
colour-magnitude diagram and by spectral type in \S\ref{disc}, which
includes a model the results.  Finally, we provide a summary in
$\S\ref{conc}$.  All magnitudes are in the AB \citep{f96} magnitude
system unless otherwise noted.  We assume an $H=$ 70, $\Omega_M=$ 0.3,
$\Omega_\Lambda=$ 0.7 cosmology.  LBG separations stated in \kpc\
refer to physical scales and those stated in \Mpc\ are in comoving
coordinates, unless otherwise noted.


\section{OBSERVATIONS}\label{obs}

The Deep fields of the CFHTLS\footnote{General information for the
CFHTLS Deep fields, such as location, cadence, and data products can
be found at:
www.cfht.hawaii.edu/Science/CFHLS/cfhtlsdeepwidefields.html and the
associated links} are used for the photometry in this work and consist
of four widely separated square-degree MegaCam pointings imaged in
five filters ($u^*g'r'i'z'$) during the years 2003 - 2008.  We combine
the highest quality data (seeing $<0.75"$ FWHM) from the first four
years (with consistent $i'$-band data) and generate deep, m$_{lim}\sim$
27, stacked images for each of the five filters.  Further details on
the data reduction and stacking process can be found in
\citet[][Supplementary Information]{cooke09b}.

Sources are detected using the {\it SExtractor} \citep{ba96} software
v.2.8.6 down to the limiting magnitude of the stacked images in each
field.  Detections in the $i'$-band images (restframe $\sim$1900\AA)
are used to define the LBG catalogues for each field.  The limiting
magnitudes are defined as the magnitudes in which we retrieve 50\% of
fake point-like (\zzz\ LBG-like) sources placed in the images.  We
compare the results per field with the number counts of real
detections per magnitude interval and find that the two methods are
consistent and that {\it SExtractor} may overestimate the limiting
magnitude when using 0.198 mag (5$\sigma$) uncertainties.  The
limiting magnitudes vary between field and filter, with $i'$-band
limiting magnitudes ranging from $i_{lim}=$ 26.4 - 26.8 mag and
$u^*g'r'$ limiting magnitudes ranging from m$_{lim}\sim$ 27.0 - 27.5
mag.  As such, we refer to the full LBG sample for the four Deep
fields as the ``$i'\lesssim$ 26.4'' sample hereafter, as this is the
limiting $i'$-band magnitude for identifying LBGs in the shallowest
field.  We note that, although other fields probe to deeper $i'$-band
magnitudes, this value is representative of our \zzz\ LBG sample
magnitude limit because of the need for deeper imaging in the $u^*$
and $g'$ filters for colour selection and for spectral-type
colour-magnitude selection as described in \S\ref{z3colors}.

Follow-up spectroscopy of CFHTLS \zzz\ LBG colour selected sources
were acquired from 24 January 2009 through 10 March 2011 using the Low
Resolution Imaging Spectrometer
\citep[LRIS;][Appendix]{oke95,steidel04} on the Keck I telescope.
These data were obtained using either the 400/3400 or the 300/5000
grism on the blue arm and the 400/8500 grating on the red arm.  Seeing
ranged from $\sim$0.6 - 1.1 arcsec, FWHM, and individual integrations
were 1200s.  Because the data were gathered in conjunction with other
research, the total exposure times per multi-object slitmask ranged
from $2400-8400$s.

We targeted LBGs from m$_{r'}\sim$ 22 - 27 and thus obtained continuum
a signal-to-noise ratios (S/N) near restframe 1700\AA\ from a S/N
$\sim$ 10 to essentially non-detection for \lya\ emitting objects.  We
note that \zzz\ LBGs can be reliably identified in continuum spectra
with a S/N of only a few from their strong UV ISM features
\citep[e.g.,][]{steidel98,steidel03,steidel04,shapley03,cooke06} and
from \lya\ emission, when present, which is detected at higher
significance.  All objects meet the \zzz\ colour-selection criteria
and the few spectra that display a single emission line but have
continua too faint to reliably identify ISM absorption features, the
emission is assumed here to be \lya.

From 178 targeted spectra, 68 have high enough continuum S/N or \lya\
S/N for confident identification.  We categorise the remaining spectra
as `unknown' as a result of their low S/N caused mainly by shortened
total slitmask integration times due to primary science programme
constraints or as a result of weather.  Of the identified spectra, two
are $z<$ 2 sources, two are \zzz\ LBGs with evidence of AGN activity,
and three are \zzz\ LBGs with evidence of double \lya\ peaks and
potentially two closely spaced continua in the 2-D spectra and two
flux peaks in the images (i.e., potential interactions).  These seven
objects were omitted from the \lya\ EW analysis.

The spectral-type criteria, spectrophotometry, and relevant tests
presented here use the larger spectroscopic dataset ($\sim$800 \zzz\
LBGs) of \citet[][hereafter S03]{steidel03} and composite spectra of
\citet{shapley03}.


\section{LYMAN BREAK GALAXY SELECTION}\label{z3colors}

We design the colour selection criteria for the CFHTLS to identify
\zzz\ LBGs over the same redshift path as S03 to aid in direct
comparison to the results of C09.  We determine the criteria using (1)
the color evolution of galaxy templates, (2) spectrophotometry using
\zzz\ LBG composite spectra, and (3) the identified LBG spectra in the
fields.

Firstly, we convolve seven star forming, one QSO, and two early-type
galaxy templates with the throughput of the $u^*g'r'i'z'$ filters,
MegaCam detector quantum efficiency, and the atmospheric extinction of
Mauna Kea and then evolve the templates from $z=$ 0--3.5 in multiple
colour-colour planes.  We vary the amount of absorption caused by
optically thick systems in the line of sight (D$_A$) and include a
star forming template that brackets 0.2--2.0 times the value measured
for average LBGs at \zzz.

Secondly, we compute the spectrophotometric colors for four \zzz\ LBG
composite spectra. \citet{shapley03} separated 794 \zzz\ LBG spectra
into quartiles based on \lya\ EW.  The composite spectra are formed
from these data and consist of $\sim$200 LBGs from each quartile.  As
such, the composite spectra reflect a consistent increase in net \lya\
EW and decrease in reddening, ISM line widths, and star formation
rates.  We randomly pull from the observed redshift and $\cal{R}$
magnitude distributions for each quartile to compute $\cal{R}$-band
fluxes for each composite spectrum.  We perform this analysis 1000
times while measuring the corresponding flux in the $U_n$ and $G$
bandpasses to determine the colors for each spectrum.

\begin{figure}
\begin{center}
\scalebox{0.37}[0.37]{\rotatebox{90}{\includegraphics{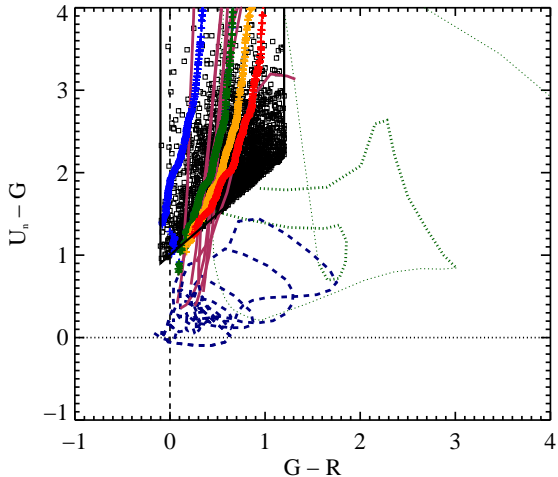}}}

\caption {\small Colour-colour plot for the data of \citet{steidel03}.
The various curves trace the colour-colour evolution of 10 galaxy
templates from $z=$ 0--3.5.  Star-forming templates and a QSO template
are indicated by the dashed (blue) curves for $z=$ 0--2.5 and the
solid (maroon) curves for $z=$ 2.5--3.5.  Dotted (green) curves trace
the evolution of early-type galaxies from $z=$ 0--1 (thick) and $z=$
1--3.5 (thin) for completeness but are less reliable beyond $z\sim$ 1.
The four tracks shown by the blue, green, yellow, and red crosses
indicate the evolution of LBG composite spectra with decreasing \lya\
EW, respectively, over the redshift path of the \citet{steidel03}
sample ($z\sim$ 2.5--3.5).  The \zzz\ LBG colour-selection region is
bounded by the thick lines. Squares denote spectroscopically confirmed
LBGs. }

\label{UnGR}
\end{center}
\end{figure}

We test our spectrophotometry in the ($U_n-G$) vs. ($G-\cal{R}$) color
plane and on the colour-magnitude diagram (CMD).  The latter is
discussed in \S\ref{types}.  These tests reveal that the composite
spectra are very representative of the average spectrum in each
quartile and thus accurately trace the colour-colour evolution and
colour-magnitude distribution of each quartile and the full population
when combined.  The colour-colour evolution tracks for the composite
spectra over the exact redshifts of the S03 survey, $\langle z\rangle
=$ 2.96 $\pm$0.26, are traced by the crosses in Figure~\ref{UnGR}.

\begin{figure}
\begin{center}
\scalebox{0.37}[0.37]{\rotatebox{90}{\includegraphics{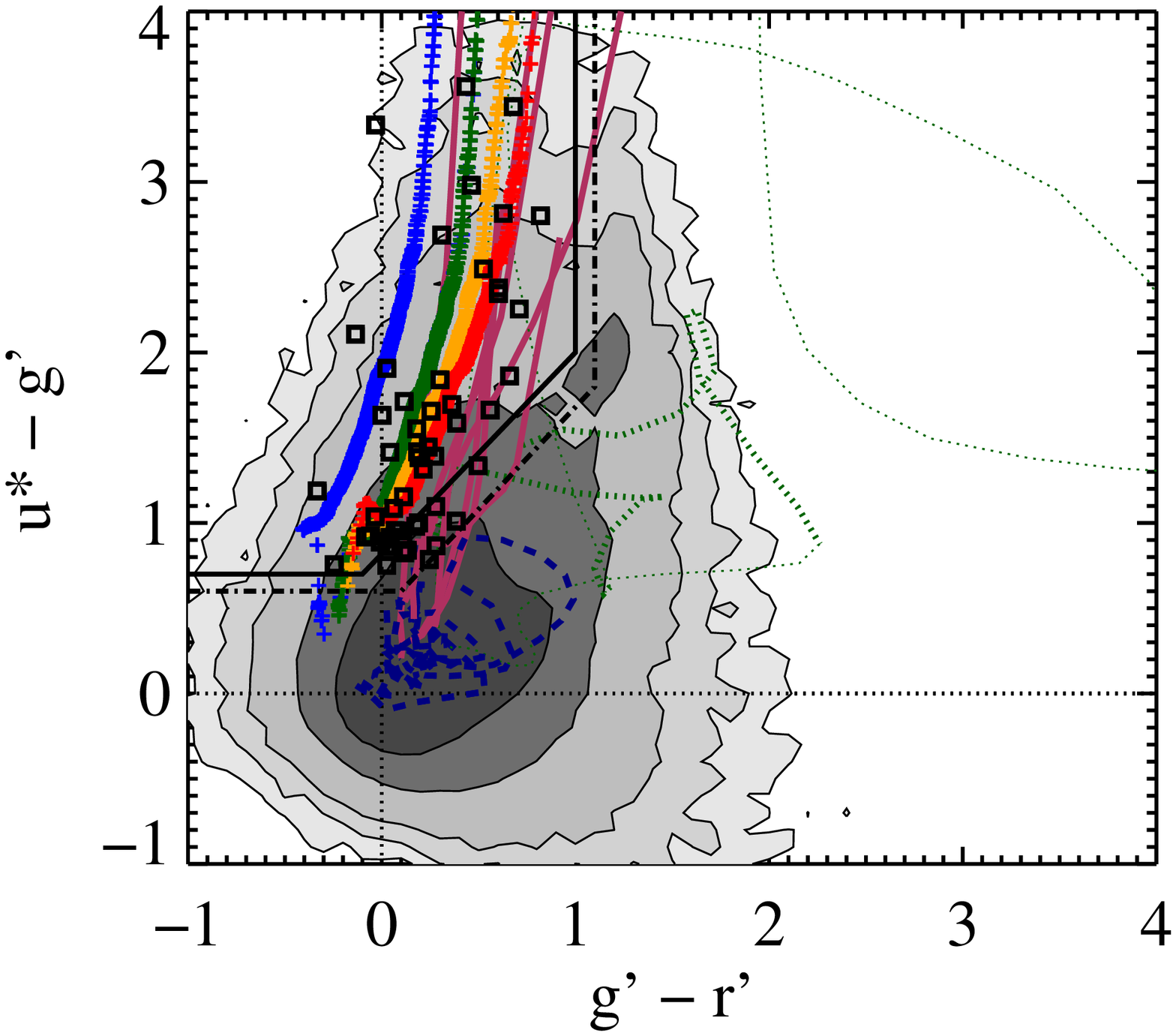}}}
\scalebox{0.37}[0.37]{\rotatebox{90}{\includegraphics{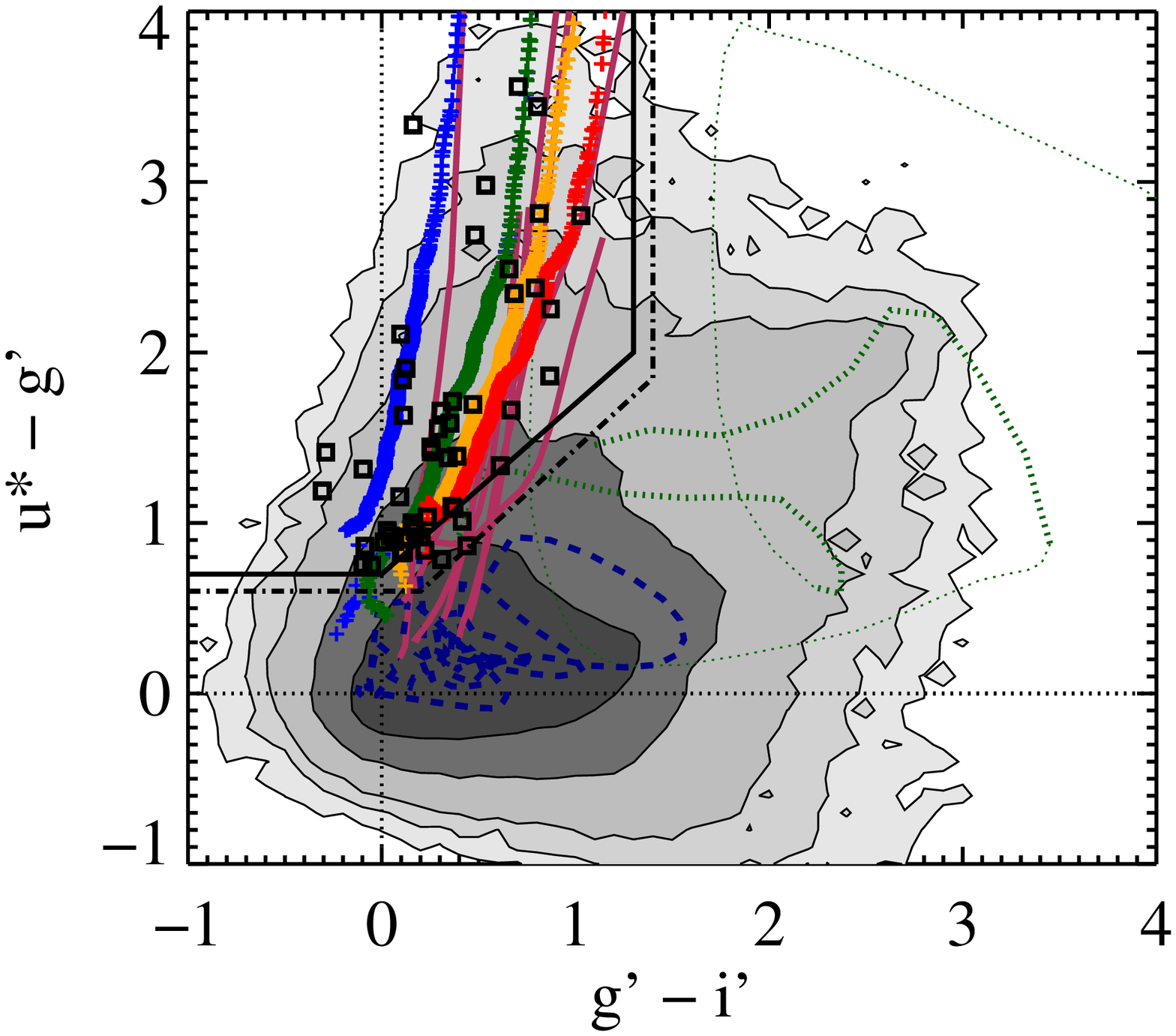}}}
\scalebox{0.37}[0.37]{\rotatebox{90}{\includegraphics{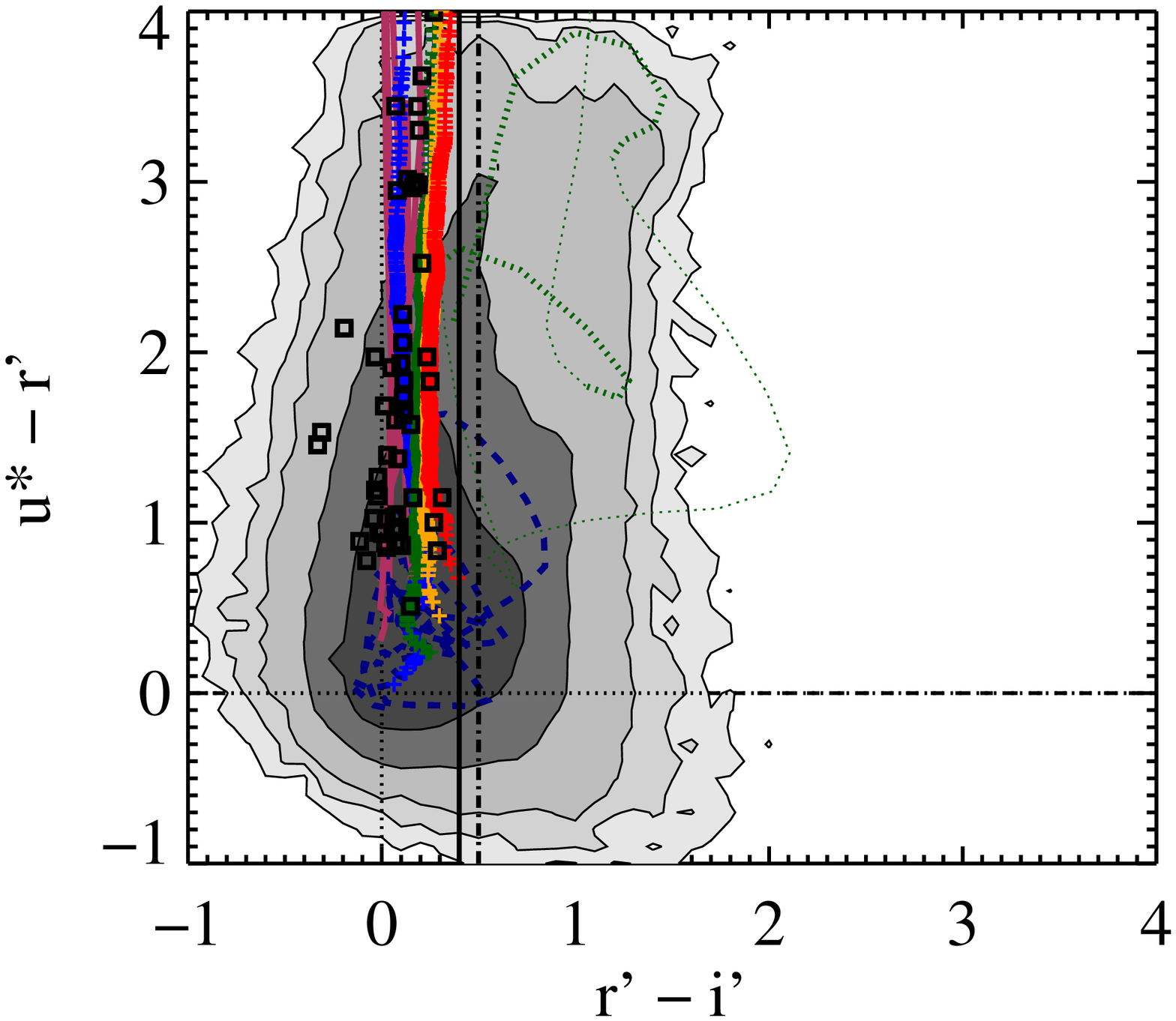}}}

\caption {\small Colour-colour plots for the CFHTLS survey Deep fields
plotted similarly to Figure~\ref{UnGR}.  Grey contours trace the
$\sim$2 million sources detected in the stacked images for the four
square-degree fields with each level reflecting a 4$\times$ greater
number density than the previous lighter shaded contour.  Black solid
lines mark the boundaries of the conservative colour selection regions
used in this work whereas the dot-dash lines mark the regions for a
more standard criteria. }

\label{colorplots}
\end{center}
\end{figure}

Confident that our spectrophotometry duplicates the $U_nG\cal{R}$
colour selection, we determine the colour evolution of the composite
spectra when passed through the CFHTLS filters.  By doing so, we are
``observing'' the S03 objects with the $u^*g'r'i'z'$ filters.  Both the
template evolution and composite spectrophotometry are studied in all
permutations of colour-colour space, however, because the $z'$-band
data is shallower than the other bands and has accompanying larger
photometric errors, we do not include these data when determining the
colour-selection criteria.  The colour evolution of the composite
spectra and the star forming templates can be seen in the three
colour-colour planes shown in Figure~\ref{colorplots}.

The general LBG colour-selection criteria shown by the dot-dash line
in Figure~\ref{colorplots} is typical of \zzz\ colour selection
regions designed to probe a similar redshift path as that of S03.
These criteria avoid the low redshift tail for some templates and
composite spectra where the density of objects in the field is high
(cf. the darkest contour in each panel of Figure~\ref{colorplots}).
Our spectra confirm that the colour-selection criteria are highly
effective and yield the same redshift distribution as S03 (see
\S\ref{redshifts}).  In an effort to improve the LBG purity of the
colour-selection criteria, we make conservative cuts (solid lines in
Figure~\ref{colorplots}) just inside the general colour-selection
regions to account for photometric uncertainties that result in
$\sim$0.1 mag scatter in the colour-colour plane and to further remove
LBG selection from the central high density region of low redshift
field objects and the regime of lower redshift reddened elliptical
galaxies and the stellar locus.

We define the following selection criteria with the aim of selecting a
clean sample of $\langle z\rangle=$ 3.0 $\pm$0.3 LBGs for the work
presented here.
\begin{equation}
(u-g) > 0.7
\end{equation}
\begin{equation}
(u-g) > 1.2*(g-r) + 0.9
\end{equation}
\begin{equation}
-1.0 < (g-r) < 1.0  
\end{equation}
\begin{equation}
(u-g) > (g-i) + 0.7
\end{equation}
\begin{equation}
-1.0 < (g-i) < 1.3  
\end{equation}
\begin{equation}
(r-i) < 0.4  
\end{equation}

Applying equations 1 - 6 to the stacked images of the four
square-degree CFHTLS fields identifies 57,382 \zzz\ LBGs for our
m$_{i'}\sim$ 26.4 sample.

To further assess the efficiency of our criteria and the make-up of
the selected populations, we analyse the follow-up Keck spectroscopy.
We find that two of the 68 spectra (3\%) are low redshift objects with
colors that mimic \zzz\ LBGs.  This low fraction helps confirm the
effectiveness of our criteria.  As mentioned in \S\ref{obs}, two
objects show signs of AGN activity and this fraction is consistent
with that found in the larger spectroscopic samples of S03 and
\citet{cooke06}.  Finally, three objects appear to be interacting
systems which is consistent with the fraction found in
\citet{cooke10}.  We conclude that our criteria is highly efficient
and produces samples that are representative of the full LBG
population.

\subsection{Lyman Break Galaxy Sub-Samples}\label{subsamples} 

\subsubsection{Colour and Magnitude}\label{cm}

As described in \S\ref{tests}, we use simple divisions of the
($g'-i'$) versus $i'$ CMD to test the effects of colour and magnitude
on the various LBG ACF sub-samples.  We split the CMD in half
horizontally, in observed ($g'-i'$), to produce samples to test colour
effects.  Similarly, we split the CMD in half vertically, in observed
$i'$, to test magnitude effects.  Below, we first detail the more
complicated process to divide the CMD into regions that contain pure
samples of LBG spectral types, those having different net \lya\ EW.
Finally, we test the effects of colour and magnitude on the LBG
spectral type samples in \S\ref{equal}.

\subsubsection{Spectral types}\label{types} 

LBGs display \lya\ in absorption, emission, or a combination of both.
The net \lya\ EW distribution for LBGs at \zzz\ has a wide range, from
net \lya\ EW $\lesssim$ -50\AA\ to $\gtrsim$ 200\AA\footnote{Using the
convention in the literature, a negative net \lya\ EW corresponds to
net \lya\ in absorption and a positive net \lya\ EW corresponds to net
\lya\ in emission.  Typically, LBGs that have net \lya\ EW near zero
exhibit both \lya\ in emission and absorption.}, with an asymmetric
peak near zero \citep{shapley03}.  As mentioned earlier, there is a
strong relationship between \lya\ EW and other spectroscopic
properties.  LBGs with net \lya\ EW in absorption show redder UV
continua (see Figure~\ref{filters}), higher star formation rates,
stronger/broader ISM absorption lines, and larger line velocity
offsets with respect to systemic redshifts as compared to LBGs with
net \lya\ EW in emission.  Thus, net \lya\ EW is a direct indicator of
multiple spectroscopic features and LBG properties that are directly
relevant to this work.

Although \lya\ is the dominant spectroscopic feature of LBGs, the
relatively low S/N of many of the spectra affects our ability to make
a precise measure of the net EW.  We find that the \lya\ forest and
absorption features near \lya\ make an accurate determination of the
continuum level difficult and result in a net \lya\ EW uncertainty of
$\sim$ 25\% for \lya\ emission features and $\sim$25 - 50\% for \lya\
absorption features.  Consequently, we treat our LBG spectra in a
similar manner as C09.  We divide the spectra into two groups with net
\lya\ EW significantly removed from net \lya\ EW $=$ 0, relative to
the uncertainties, to classify LBGs with dominant \lya\ in absorption,
termed `aLBGs' and \lya\ in emission, termed `eLBGs'.  We adopt net
\lya\ EW $<$ -10\AA\ for aLBGs and net \lya\ EW $>$ 20\AA\ for eLBGs
based on the range of net \lya\ EW for quartile 1 LBGs (strongest net
\lya\ EW in absorption) and quartile 4 LBGs (strongest net \lya\ EW in
emission) of \citet{shapley03} and from similar net \lya\ EW results
of our spectroscopic sample.  All other LBGs are classified as ``grey
area'' LBGs, or `gLBGs', with net \lya\ EW near zero.  As a note, the
\lya\ EW cut places most eLBGs under conventional definitions of \lya\
emitters (LAEs) detectable in deep narrow-band surveys.

\subsubsection{Spectral type photometric selection criteria}
\label{photcrit}

C09 identifies a natural segregation of the aLBG and eLBG net \lya\ EW
distributions on the CMD and uses that property to isolate highly pure
samples of the two sub-populations.  The criteria were determined
using the S03 data set which contains $\sim$800 $U_nG\cal{R}$-selected
spectra.  The spectral type selection technique exploits the inverse
relationship between the the UV continuum near $\sim$1700\AA\ and the
combination of continuum, \lya\ feature, and \lya\ forest near
$\sim$1200\AA.  As a result, using broadband information alone,
$>95$\% pure samples of each LBG spectral type can be confidently
isolated.

The four-year stacked images of the CFHTLS Deep fields enable LBG
detections over $\sim$$10\times$ the area and $\sim$1--1.5 mags deeper
than the S03 survey data considered in the C09 analysis.  The CFHTLS
data provide the necessary large samples of the LBG spectral types to
perform the first detailed study of their spatial distribution.
However, to properly apply the results of C09 to the data here, we
need to correct for the differences between the MegaCam and S03
filters.

\begin{center}
\begin{figure}
\scalebox{0.4}[0.4]{\rotatebox{90}{\includegraphics{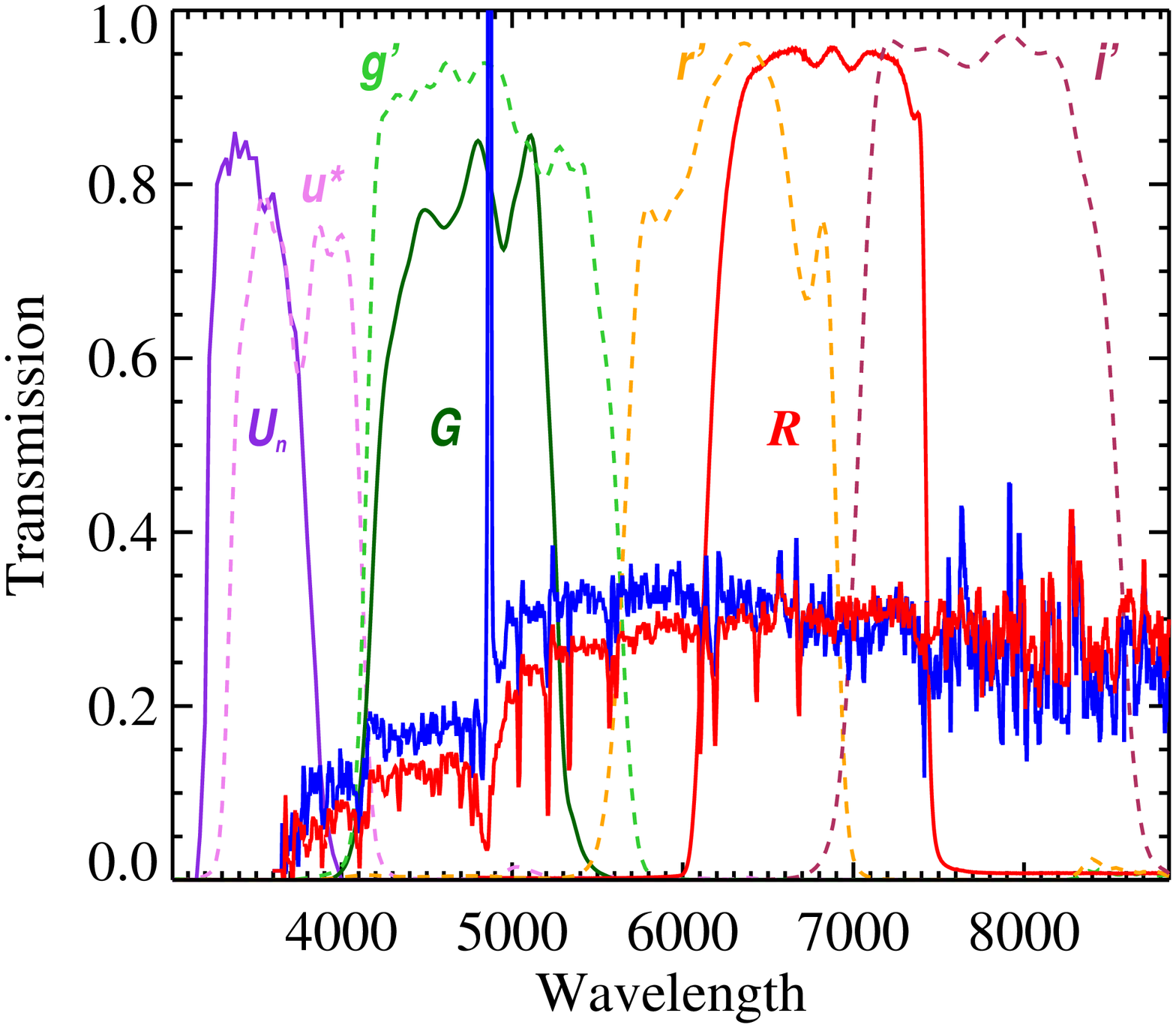}}}

\caption {\small Lyman break galaxy (LBG) spectral regions sampled by
the filters studied here.  Solid curves show the \citet{steidel03}
$U_n,G,\cal{R}$ filters and the dashed curves show the CFHTLS
$u^*,g',r',$ and $i'$ filters.  Overlaid are composite \zzz\ LBG
galaxy spectra shifted to $z=$ 3; an aLBG (dominant \lya\ in
absorption) in red and an eLBG (dominant \lya\ in emission) in blue.
Although $\cal{R}<$25.5 LBGs with dominant \lya\ in absorption are
$\sim$0.4 mag more luminous on average \citep{shapley03}, the
templates shown here are normalised in the $\cal{R}$ filter to help
illustrate the flux differences between the $G$ and $\cal{R}$ filters
that segregate the spectral types on a colour-magnitude diagram.  The
$g'$ and $i'$ filters produce similar flux differences.}

\label{filters}
\end{figure}
\end{center}

The relevant filters are shown in Figure~\ref{filters}.  The
sensitivities for the CFHT $g'$ filter (4872/1455; central
wavelength/bandwidth in \AA) and S03 $G$ filter (4780/1100) are
similar, with the $g'$ filter being somewhat broader and redder.  The
S03 $\cal{R}$ filter sensitivity (6830/1250) falls between those of
the CFHT $r'$ (6282/1219) and $i'$ (7776/1508) filters.  Because LBG
continua are relatively flat over the wavelength ranges probed by the
$r'$, $\cal{R}$, and $i'$ filters, and because of the similarity
between the $g'$ and $G$ filters, we expect the corrections to the
criteria used in C09 to be relatively small.  We quantify the
corrections using a spectrophotometric analysis and by using the
distributions of our Keck CFHTLS spectra.

The spectrophotometry of the LBG composite spectra as described above
accurately reproduces the magnitude and colour means and dispersions
on the $G$ vs. $G-\cal{R}$ CMD for each of the four quartiles
(Table~\ref{S03means}), as well as the full CMD distribution of S03
when combined.  The exception is quartile 4 containing the strongest
\lya\ emission (eLBGs) which has an offset in the colour mean by -0.12
mag.  The contribution to the average \lya\ EW from a small number of
strong \lya\ emitters results in a bias of the composite spectrum
colour as compared to the entire quartile sample.  Because we do not
have access to the individual spectra, we were not able to directly
correct for this effect.  Instead we applied a +0.1 mag correction to
the $g'$-band values of the composite spectrum to counter the bias.

Regarding the correction, it is important to note three points: (1)
the correction is small, (2) there is no effect on the magnitude mean
or dispersion ($i'$-band based), and (3) without the correction the
eLBG mean would move in a direction away from the aLBG mean.  As can
be seen below, the correction provides a more conservative estimate of
the true aLBG and eLBG colour distribution separations.  This is
because the selection of the spectral types is based on a fixed
separation from the distribution means.  Because the correction moves
the means of the two distributions closer, the fixed separation probes
further from the respective spectral type mean, resulting in purer
spectral-type samples at the cost of reducing the total number of
objects.  We conclude that, although the colors and magnitudes of
individual spectra vary within each quartile, the composite spectra
can be used to compute net \lya\ EW means and dispersions on the CMD
for the purposes here in lieu of individual spectra.

\begin{figure}
\begin{center}
\scalebox{0.4}[0.4]{\rotatebox{90}{\includegraphics{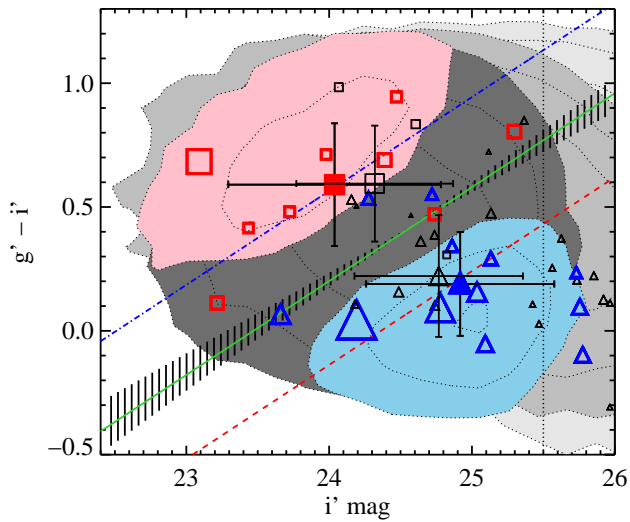}}}

\caption {\small Colour-magnitude diagram of \zzz\ Lyman break galaxy
(LBG) spectral type distributions in the CFHTLS Deep fields.  Shown in
contours are the {\it SExtractor} 5$\sigma$ detections (light grey),
the $i'\lesssim$ 26.4 LBG sample (medium grey), and the composite
spectra matched to the survey depth (m$_{\cal{R}}\lesssim$ 25.5;
vertical dotted line) of \citet[][dark grey]{steidel03}.  Each inner
contour ridge represents 4$\times$ the number density of the previous.
CFHTLS spectroscopically confirmed LBGs are shown as open squares and
triangles with the symbol size reflecting the relative net \lya\ EW
strength with respect to zero such that larger squares have stronger
absorption and larger triangles have stronger emission.  Coloured
symbols are objects that meet our aLBG/eLBG \lya\ EW criteria whereas
smaller black symbols are consistent with zero net \lya\ EW.  The
error crosses with solid symbols mark the respective (m$_{i'}\lesssim$
25.5) means and 1$\sigma$ dispersions of the spectra, whereas the open
symbols denote the spectrophotometric means and 1$\sigma$ dispersions.
The solid (green) line divides the two spectrophotometric
distributions and the dashed (red) and dot-dashed (blue) lines denote
$\gtrsim$2.5$\sigma$ from either distribution mean.  The hatched
region denotes the range of slopes from the spectra and the
spectrophotometric analysis.  The pink and light blue regions
represent spectrophotometric $U_nG{\cal R}$ aLBGs and eLBGs,
respectively, meeting the spectral-type criteria in \citet{cooke09a}
transformed into the $g'$ and $i'$ colour-magnitude space. }

\label{cuts}
\end{center}
\end{figure}

\begin{figure}
\begin{center}
\scalebox{0.4}[0.4]{\rotatebox{90}{\includegraphics{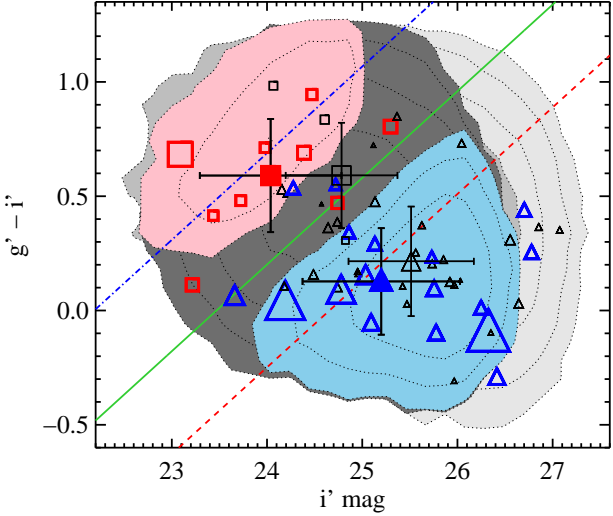}}}

\caption {\small Similar to Figure~\ref{cuts}, but reflecting the
$i'\lesssim$ 26.4 spectrophotometric analysis to match our higher
significance sample (medium grey).  The blue dot-dashed and red dashed
lines define the conservative spectra-type cuts used here {\it (see
text)}.  The cuts provide highly pure samples of each spectral type
such that $\gtrsim$95\% of eLBGs are below the blue dot-dashed line
and $\gtrsim$95\% of aLBGs are above the red dashed line.  }

\label{264cuts}
\end{center}
\end{figure}

We then use the composite spectra to `observe' the LBGs of S03 with
the MegaCam filters.  We use the redshift and $\cal{R}$ magnitude
distributions of the S03 data to compute the $i'$ magnitude and
($g'-i'$) colour distributions for each LBG when passing the composite
spectra through the $g'$ and $i'$ filters.  We do this for the
magnitude range of the S03 data and extend this $\sim$1 magnitude
fainter to estimate the values for the full CFHTLS sample.  The
results are listed in Table~\ref{CFHTmeans} and shown in
Figures~\ref{cuts} and~\ref{264cuts}.  The composite spectra do a good
job in duplicating the overall form of the distributions on the $i'$
vs. ($g'-i'$) CMDs.  The broader form of the distributions in the
bluer regions of the CMDs, i.e., the small extension of bright, blue
LBGs, is nearly identical to the composite spectra distribution on the
$G$ vs. $G-\cal{R}$ CMD and the tests with the S03 sample informs us
that the CFHTLS means and dispersions are similarly accurate.

Next, we compute the ($g'-i'$) and $i'$ aLBG and eLBG means and
dispersions for our confirmed Keck spectra.  We determine the values
with $i'\lesssim$ 25.5 to compare directly with the S03 data and those
for the $i'\lesssim$ 26.4 sample.  The results are listed in
Table~\ref{CFHTmeans} and illustrated in Figures~\ref{cuts}
and~\ref{264cuts}.  The means and dispersions of the $i'\lesssim$ 25.5
spectra and spectrophotometry are consistent, supporting the analysis
of the LBG composite spectra with the CFHTLS filters.

We repeated this analysis for \zzz\ LBG ($g' - r'$) colour and $r'$
magnitude distributions.  The results of this investigation show that,
as expected from the wavelengths probed by each of the filters, the
aLBG and eLBG ($g' - r'$) and $r'$ distributions are closer together
on the CMD and have more overlap than distributions using $G$ and
$\cal{R}$ or $g'$ and $i'$ filters.  LBG spectral types are separated
in part by the slope of their continua longward of restframe
$\sim$$1500$\AA, with increasing differences with increasing
wavelength (cf. Figure~\ref{filters}).  As such, we find that the
larger differences provided by the $g'$ and $i'$ filters are more
effective in separating the distributions on the CMD as compared to
the $g'$ and $r'$ filters for the redshift path probed.

For the $i'\lesssim$ 25.5 sample, we follow the spectral-type approach
of C09 and define a primary cut that statistically divides the two
distributions (solid green line in Figure~\ref{cuts}).  The aLBG
region is then determined as the area to the upper left of (brighter
and redder than) a line placed 1.5$\sigma$ redward of the primary cut,
away from the eLBG distribution mean, with the same slope.  Similarly,
the eLBG spectral-type region is the area to the lower right of
(fainter and bluer than) a line placed $\sim$1.5$\sigma$ away from the
primary cut and blueward of the aLBG distribution mean.  As a result,
each spectral type region is $\gtrsim$2.5$\sigma$ (2.5$\sigma$ at its
closest) from the other spectral type distribution mean.  The number
of aLBG and eLBG Keck spectra is relatively small to accurately
determine the slope of the primary cut alone but yield means and
dispersions similar to the spectrophotometric values.  Because the
position and slope of the primary cut from the spectrophotometric
analysis is similar to that determined by the spectra (hatched region
in Figure~\ref{cuts}), we use the average of the two values.

Redshift identifications and \lya\ EW measurements of faint,
$i'\gtrsim$ 25.5 spectra can only be efficiently determined for LBGs
with \lya\ in emission, therefore we only estimate the $i'\gtrsim$
25.5 eLBG distribution.  Although we have identified objects with
dominant \lya\ emission to $i'\sim$ 27, interestingly, we find none in
the region bounded by $i'\gtrsim$ 25.5 and $(g'-i')>$ 0.5.  Given that
LBGs meeting the colour-selection criteria are detected with
$i'\gtrsim$ 25.5 and $(g'-i')>$ 0.5, the region contains either (1)
aLBGs with a similar level of purity as the $i'\lesssim$ 25.5 sample
(i.e., no change in color with magnitude), (2) LBGs with net \lya\ EW
$\sim$ 0 (i.e., gLBGs) only, or (3) a combination of the two.  The
$i'\lesssim$ 26.4 spectrophotometric analysis makes no assumptions of
a colour trend for $i'\sim$ 25.5--26.4 objects and therefore the
distributions differ from the $i'\lesssim$ 25.5 distributions in
magnitude only.  As a result, the spectrophotometric aLBG analysis
provides an estimate of scenario (1).  Because we are not able to
confidently identify our $i'\gtrsim$ 25.5 spectra as aLBGs, and the
lack of eLBGs, results in the $i'\lesssim$ 26.4 aLBG mean being
unaffected from the $i'\lesssim$ 25.5 value, thus providing an
estimate of scenario (2).  As a result, the two colour and magnitude
mean and distribution estimates bracket the range of values for the
$i'\lesssim$ 26.4 aLBG sample for all three scenarios.

\begin{table}
\begin{center}
\caption\normalsize{Colour and magnitude means and dispersions}
\label{S03means}
\begin{tabular}{lccccc}
\hline
Spectral type$^{a}$ & $G-\cal{R}$ & $G-\cal{R}$ & 
$\cal{R}$ mag & $\cal{R}$ mag & $\cal{R}$ mag \\
 & mean & 1$\sigma$ & mean & 1$\sigma$ & limit$^{b}$ \\
\hline
S03 q1 data & 0.75 & 0.25 & 24.44 & 0.53 & 25.5\\
S03 q1 composite & 0.77 & 0.25 & 24.44 & 0.55 & 25.5\\
S03 q2 data & 0.68 & 0.26 & 24.51 & 0.50 & 25.5\\
S03 q2 composite & 0.70 & 0.25 & 24.52 & 0.52 & 25.5\\
S03 q3 data & 0.60 & 0.25 & 24.68 & 0.52 & 25.5\\
S03 q3 composite & 0.60 & 0.25 & 24.68 & 0.53 & 25.5\\
S03 q4 data & 0.45 & 0.30 & 24.84 & 0.58 & 25.5\\
S03 q4 composite & 0.33 & 0.29 & 24.85 & 0.59 & 25.5\\
\hline
\end{tabular}
\end{center}
$^{a}${q1 - q4 are abbreviations for quartiles 1 - 4 of
\citet{shapley03}} 
$^{b}${One field (of 17) has a limiting magnitude of $\cal{R}\sim$ 26.0
and is accounted for in the composite spectrum analysis.}  
\end{table}

Increasing the 1.5$\sigma$ displacement from the $i'\lesssim$ 25.5
primary cut to 2.0$\sigma$ is expected to produce pure $i'\lesssim$
26.4 samples while considering all three scenarios and the uncertainty
of the full census of $i'\gtrsim$ 25.5 LBGs.  For aLBGs, the increase
to 2.0$\sigma$ avoids including gLBGs and the tail of the eLBG
distribution but sacrifices the total number of aLBGs.  For eLBGs, a
2.0$\sigma$ displacement similarly helps to omit the far tail of the
aLBG distribution and simply avoids the $i'\gtrsim$ 25.5 and
$(g'-i')>$ 0.5 region over the extent of our $i'\lesssim$ 26.4 sample.

The choice of a 2.0$\sigma$ cut comes at the cost of the total number
of aLBGs and eLBGs used for our correlation function analysis, but the
large numbers available from the four CFHTLS fields gives us the
option to attack this problem conservatively.  We vary the spectral
type cut parameters (slope and displacement) over a practical range
and find that there is no significant change in the overall behavior
of the correlation functions of the two spectral types.  Thus, the
main results of this paper are insensitive to moderate departures from
the spectral-type criteria defined below.

We define the $i'\lesssim$ 26.4 sample spectral-type criteria as
\begin{equation}
\mbox{aLBGs: } ~(g'- i') \ge 0.38\cdot i' - 8.9 + 2.0\sigma_E
\label{aLBG} 
\end{equation} 
\begin{equation} 
\mbox{eLBGs: } ~(g'- i') \le 0.38\cdot i' - 8.9 - 2.0\sigma_A 
\label{eLBG} 
\end{equation}

\noindent where $\sigma_E$=0.25 and $\sigma_A$=0.23 and refer to the
colour dispersions for the eLBG and aLBG distributions, respectively.
Note that the eLBG distribution ($\sigma_E$) is used to determine the
aLBG spectral-type cut and vice-versa.  In \S\ref{tests}, we present
the results from tests of other LBG sub-samples that provide insight
into the effects caused by the choice of more general slope and sample
criteria and the dependence of the correlation function on colour and
magnitude.

Applying the spectral-type criteria to the $i'<$ 26.4 sample in the
four CFHTLS Deep fields produces 9648 aLBGs and 11567 eLBGs.  Objects
in the aLBG region reside $\sim2\sigma$ from the primary cut and
$\gtrsim$3.0$\sigma$ from the eLBG distribution mean and vice-versa
for the objects in the eLBG region.  The spectrophotometric analysis
finds 3\% eLBG contamination in the aLBG sample and 1\% aLBG
contamination in the eLBG sample.  Moreover, there is zero
contamination of the Keck spectra in either samples.

All LBGs that do not meet these criteria, i.e., those in-between the
two cuts forming a swath through the middle of the CMD, are classified
as gLBGs, formally defined as

\begin{equation} 
\begin{split}
\mbox{gLBGs: } ~(g'- i') \le 0.38\cdot i' - 8.9 + 2.0\sigma_E\\
\mbox{and}~(g'- i') \ge 0.38\cdot i' - 8.9 - 2.0\sigma_A
\label{gLBG} 
\end{split}
\end{equation}

\noindent where $\sigma_E$ and $\sigma_A$ are as defined in
equations~\ref{aLBG} and~\ref{eLBG}.  These objects are comprised of a
blend of aLBGs and eLBGs, with a large fraction consisting of LBGs
with net \lya\ EW $ \sim$ 0.  We also study this population for
completeness and for added insight into the behaviour of the aLBG and
eLBG correlation functions.  Finally, we note that no
spectroscopically confirmed aLBGs are found in either the $(g'-i')$
versus $i'$ eLBG region or the equivalent ($G - \cal{R}$) versus
$\cal{R}$ eLBG region using a 2.0$\sigma$ cut for the CFHTLS Keck
spectra or the larger S03 spectroscopic sample.  These highly pure
eLBG samples reinforce the use of simple broadband criteria as an
efficient means to amass large numbers of \zzz\ LAEs and \lya\
absorbers (LAAs) quickly and inexpensively, relative to conventional
narrow-band or blind spectroscopic surveys.

\begin{table}
\begin{center}
\caption\normalsize{CFHTLS colour and magnitude means and dispersions}
\label{CFHTmeans}
\begin{tabular}{lccccc}
\hline 
   Type   & $g'-i'$ &  $g'-i'$  & $i'$ mag & $i'$ mag  & $i'$ mag\\
 &  mean   & 1$\sigma$ &   mean   & 1$\sigma$ &limit$^{a}$\\
\hline
aLBG data &  0.59  &   0.25   &  24.04  &   0.74   & 25.5\\
aLBG sim. &  0.59  &   0.23   &  24.32  &   0.55   & 25.5\\
eLBG data &  0.19  &   0.21   &  24.92  &   0.66   & 25.5\\
eLBG sim. &  0.22  &   0.24   &  24.76  &   0.59   & 25.5\\
\hline
aLBG data &  0.59  &   0.25   &  24.04  &   0.74   & 26.4\\
aLBG sim. &  0.59  &   0.23   &  24.78  &   0.58   & 26.4\\
eLBG data &  0.13  &   0.23   &  25.20  &   0.83   & 26.4\\
eLBG sim. &  0.20  &   0.24   &  25.49  &   0.66   & 26.4\\
\hline
\hline
\end{tabular}
\end{center}
$^{a}${Approximate {\it (see text)}.  The magnitude limit
of the CFHTLS $i'\lesssim$ 26.4 sample is only relevant to
eLBGs.}
\end{table}

\subsubsection{Redshift Distributions}\label{redshifts}

\begin{figure}
\begin{center}
\scalebox{0.45}[0.45]{\rotatebox{0}{\includegraphics{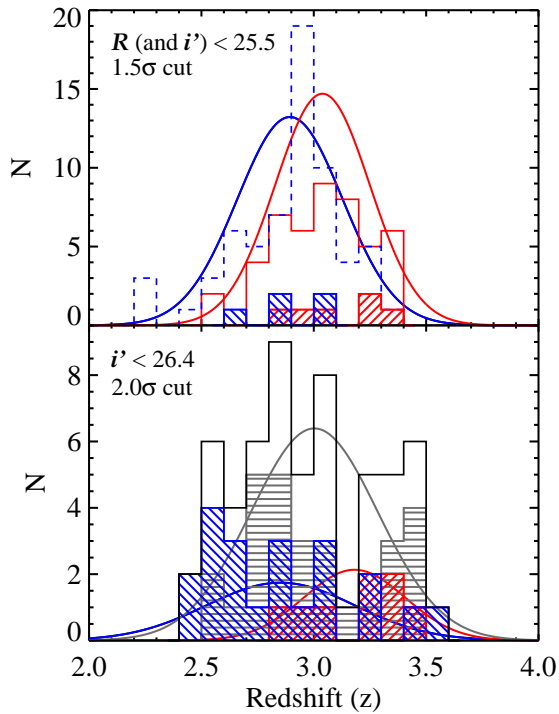}}}

\caption {\small Lyman break galaxy spectral type redshift
distributions.  {\bf Upper panel:} The dashed (blue) and solid (red)
empty histograms indicate the eLBG and aLBG redshift distributions,
respectively, for the $U_nG\cal{R}$-selected, $\cal{R}\lesssim$ 25.5,
data set of \citet{steidel03} as analysed in \citet{cooke09a}.
Overlaid are the eLBG and aLBG redshifts of the CFHTLS
$u^*g'r'i'$-selected, $i'\lesssim$ 25.5 spectra denoted by the
backward hatch (blue) and forward hatch (red) histograms,
respectively.  {\bf Lower panel:} The redshift distributions for all
spectra (no net \lya\ EW restrictions) that meet the $i'\lesssim$ 26.4
aLBG (red forward hatch), eLBG (blue backward hatch), and gLBG (grey
horizontal hatch) spectral-type selection.  In each panel, Gaussian
fits to the redshift distributions normalised to the relative total
number of objects in each sample are shown by the solid curves.}

\label{zhist}
\end{center}
\end{figure}

As discussed above, our CFHTLS \zzz\ LBG colour selection criteria was
designed to probe the same redshift path as the S03 $U_nG\cal{R}$
colour-selection criteria.  S03 reports $\langle z\rangle$ = 2.96,
$\sigma$ = 0.29 and we find $\langle z\rangle$ = 2.99, $\sigma$ = 0.28
for our $i'\lesssim$ 25.5 spectra and $\langle z\rangle$ = 2.97,
$\sigma$ = 0.31 for the full sample.  Similar to the C09 results, we
find a difference in the aLBG and eLBG redshift distributions as a
consequence of the separation of the two samples on the CMD.  The
difference occurs because higher redshift objects produce larger ($g'$
- $i'$) values and a standard candle is fainter by $\sim$0.6 mag when
redshifted from $z=$ 2.5 to $z=$ 3.5.  However the situation becomes
more complicated as aLBGs are offset in colour (redder) as compared to
eLBGs for a given redshift and the values have considerable scatter.
In C09, we find redshift distributions $\langle z\rangle=3.05,
\sigma=0.25$ and $\langle z\rangle=2.88, \sigma=0.24$, respectively,
for the S03 aLBGs and eLBGs used in that analysis.  Only a few of our
Keck spectra meet the $i'\lesssim$ 25.5 aLBG and eLBG criteria to
estimate the CFHTLS redshift distributions, but the data appear to
have a similar behaviour with $\langle z\rangle=3.05, \sigma=0.18$
(aLBG) and $\langle z\rangle=2.82, \sigma=0.23$ (eLBG) for the
1.5$\sigma$ cut.  The redshift distributions for the C09 analysis and
the CFHTLS data are shown in the upper panel of Figure~\ref{zhist}.

We find similar distributions for the $i'\sim$ 26.4 spectra meeting
aLBG and eLBG criteria using the 2$\sigma$ cuts.  However, more
relevant to this work are the redshift distributions of all objects in
the aLBG and eLBG regions, i.e., those with no net \lya\ EW
constraints, since all objects in these regions are used to compute
the spectral-type correlation functions.  From the Keck spectra, we
find $\langle z\rangle=3.18, \sigma=0.23$ for all objects in the aLBG
region regardless of spectral type and $\langle z\rangle=2.86,
\sigma=0.33$ for all objects in the eLBG region.  In addition, we find
a redshift distribution $\langle z\rangle=3.00, \sigma=0.28$ for the
gLBG sample.  Redshift histograms for the three sub-samples are shown
in the lower panel of Figure~\ref{zhist}.  Overlaid are Gaussian fits
to the distributions normalised to the total number of objects in each
sample.

Although the two spectral-type samples have a mean redshift offsets,
they have significant redshift overlap, important to the
cross-correlation function results.  Fitting Gaussian functions to the
two distributions in C09 finds $\sim$73\% overlap and similar overall
redshift ranges.  A similar result is found for the small number of
CFHTLS $i'\lesssim$ 25.5 spectra.

The fainter spectra in the CFHTLS $i\lesssim$ 26.4 sample favor
confirmation of eLBGs given the observational program constraints
(\S\ref{obs}).  Gaussian fits to the spectra in hand suggest a
$\sim$53\% redshift distribution overlap.  The overlapping redshift
path appears to be largely dictated by the aLBG redshift range,
roughly 2.5 $\lesssim z \lesssim$ 3.8.  It is important to note that
poorer representation for a given redshift, i.e., the tails of the
distributions, results only in noisier data but does not affect the
amplitude of the cross-correlation function for a given $\Delta z$.

We note that the spectra from different populations need only probe
the same redshift paths for the cross-correlation function to be
representative of the commonality of their spatial distribution.  The
random catalogs in the correlation functions help to minimise the
effect of projected pairs and the random projections of similar-sized
clustered regions should introduce a similar bias on all separation
scales.  The redshift path probed by our colour selection
($\sim$2.5--3.5) secures that the clustering scales are the same.  The
four square-degree fields of the CFHTLS include a large number of LBG
clustered regions to evenly distribute clustered regions on all
scales.  We explore the effect of redshift on the correlation
functions in more detail in \S\ref{tests}.

\section{CORRELATION FUNCTIONS}\label{results}

We compute the correlation functions on a field-by-field basis using
the auto-correlation function estimator $\omega (\theta) = (DD - 2DR +
RR) / RR$ \citep{landy93} and the corresponding cross-correlation
estimator $\omega_{1,2} (\theta) = (D_1D_2 - D_1R_2 -D_2R_1 + R_1R_2)
/ R_1R_2$, where $DD$, $DR$, and $RR$ are the data-data, data-random,
and random-random galaxy separations catalogues and the subscripts in
the cross-correlation estimator refer to the two sub-samples.  Random
catalogues are constructed to match the field dimensions probed by the
data with bright stars masked out and number densities several times
the observed values and normalised.  The correlation functions are
determined from the average of $100$ realizations and the
uncertainties are determined using $100$ jackknife error realizations,
each omitting a different $1/100$th the field area.  We determine the
integral constraint, IC, using the approach detailed in \citet{lee06}
and apply a value of IC = 0.012 to the data.  The final results
average the values for the four CFHTLS fields.  Finally, we note that
the square-degree fields of the CFHTLS probe well beyond the \zzz\ LBG
clustering correlation length ($\sim$4 \Mpc) and the multiple fields
help to minimise the effect of cosmic variance.

\begin{figure}
\begin{center}
\scalebox{0.38}[0.38]{\rotatebox{90}{\includegraphics{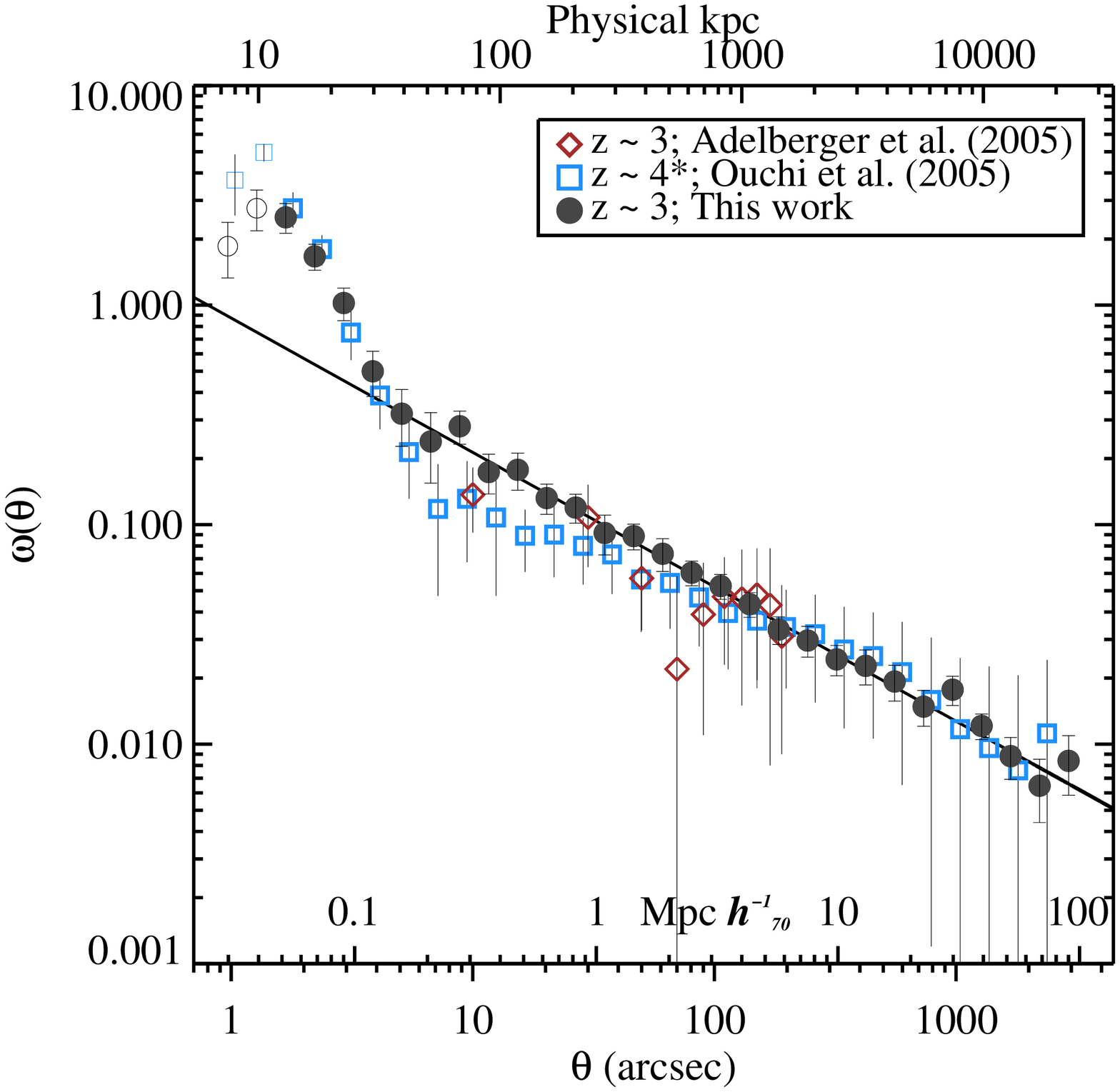}}}

\caption {\small LBG angular correlation functions (ACFs).  The CFHTLS
\zzz\ LBG ACF from this work is shown as filled circles.  For
comparison, the \zzz\ LBG ACF from \citet{adelberger05} and $z\sim4$
LBG ACF from \citet[][scaled to $z\sim3$]{ouchi05} are overlaid.  The
solid line is a power law fit to the CFHTLS ACF between $\sim$1--20
\Mpc\ separations, comoving.  The ACF shows a departure from a power
law at small scales ($\sim$0.12 \Mpc, comoving), a regime that probes
individual dark matter haloes (the one-halo term).  The two smallest
bins for our data (hollow circles) and the data of \citet{ouchi05}
(hollow squares) are potentially subject to image deblending effects.}

\label{lbgs} 
\end{center} 
\end{figure}

Figure~\ref{lbgs} presents the auto-correlation function (ACF) for the
full CFHTLS \zzz\ LBG sample.  The \zzz\ LBG ACF of
\citet{adelberger05} derived from the $17$ smaller fields of S03 and
the $z\sim4$ results of \citet{ouchi05} are overlaid for comparison.
We fit a power law of the form $\omega(\theta) = A \theta^\gamma$ to
the well-sampled two-halo regime of the ACF from $\sim$1--20 \Mpc,
yielding $\gamma=$ -0.612 and consistent with values given in the
literature.  The ACF departs monotonically from a power law at
$\sim$0.12 \Mpc, comoving, similar to that found at $z\sim4$ by
\citet{ouchi05} probing galaxies with similar luminosities and over
similar scales.  The departure occurs near the viral radius for
$\sim$10$^{11} M_\odot$ dark matter haloes at \zzz\ and is interpreted
to be caused by multiple luminous sub-halo galaxies within the parent
dark matter haloes and/or an effect of galaxy luminosity enhancement
as a result of interactions \citep{ouchi05,berrier12}.

\subsection{Spectral Type Correlation Functions}\label{spec_types}

\begin{figure}
\begin{center}
\scalebox{0.38}[0.38]{\rotatebox{90}{\includegraphics{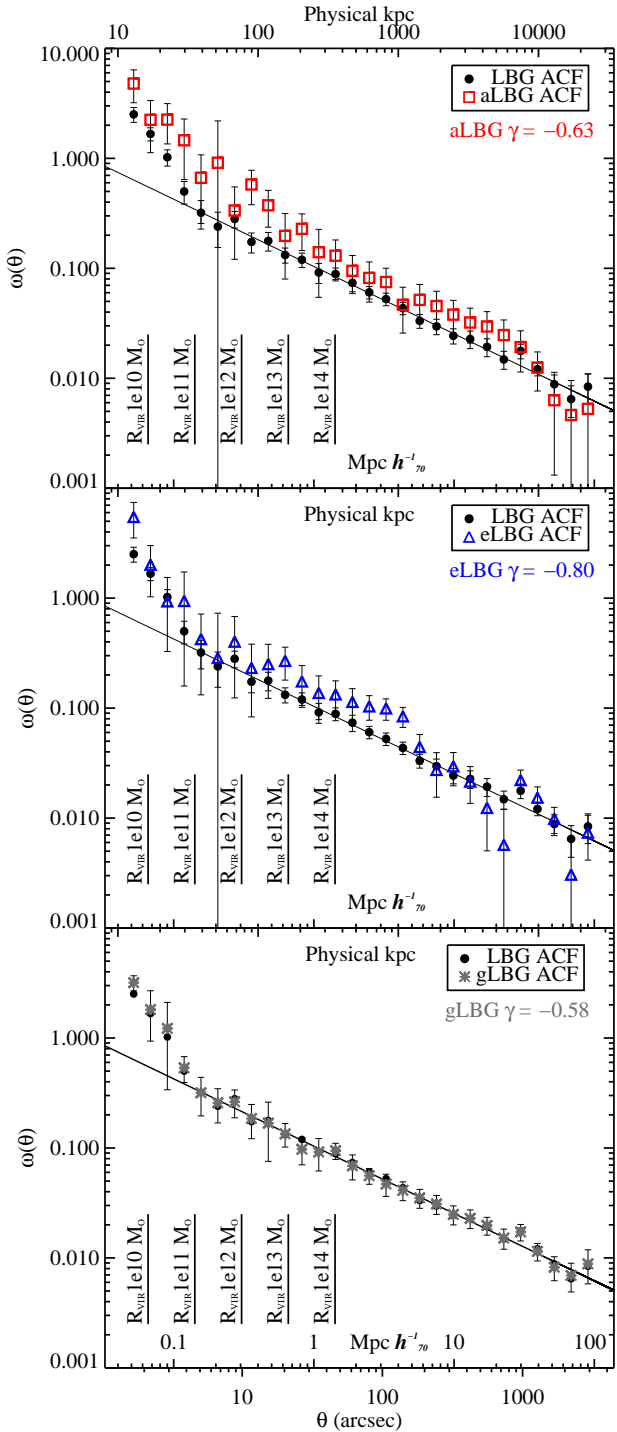}}}

\caption {\small Auto-correlation functions (ACFs) of CFHTLS Deep
field LBGs spectral-type sub-samples.  The ACF for the full LBG sample
(filled circles) is shown in each plot for comparison.  The solid line
denotes the power law fit to the full LBG ACF (see text and
Figure~\ref{lbgs}).  The virial radii, R$_{VIR}$, for
$10^{10-14}M_\odot$ haloes are indicated by the short vertical lines
and are labelled accordingly.  The slope ($\gamma$) for the power law
fit to the spectral type ACFs from $\sim$1--20 \Mpc\ is denoted below
each legend.  {\bf Upper panel:} The aLBG ACF departs from a power law
near $\sim$1 \Mpc, comoving, and displays a higher amplitude than the
full LBG population on all scales.  {\bf Center panel:} The eLBG ACF
behaves similar to the full LBG ACF and shows a potential higher
amplitude on intermediate scales ($\sim$0.5--5 /Mpc).  {\bf Bottom
panel:} The gLBG ACF closely follows the full LBG ACF.  }

\label{aelbgs} 
\end{center} 
\end{figure}

In this section, we present the ACFs and cross-correlation functions
(CCFs) for the spectral-type subsets.  The aLBG, eLBG, and gLBG ACFs
are computed as described above and shown in Figure~\ref{aelbgs} along
with the full LBG ACF for comparison.  We omit the two smallest bins
where close galaxy pairs can be difficult to separate as a result of
the seeing and {\it SExtractor} deconvolution.  In addition, we
compute the virial radii of $10^{10-14} M_\odot$ dark matter haloes
using R$_{VIR} = [(G \cdot M_{VIR}) / (100 \cdot H^2(z)]^{1/3}$
\citep[e.g.,][]{ferguson04} and plot the values on Figure~\ref{aelbgs}
for reference.  Here, we only point out the salient features and
provide a more extended examination of all ACFs features in
\S\ref{disc}.  We defer a more detailed analysis of the individual
correlation functions to a future paper.

The two main features of the aLBG ACF that stand out from the full LBG
ACF is stronger one-halo term amplitude that extends to $\sim$1 \Mpc,
comoving, ($\sim$200 kpc, physical) and the higher clustering
amplitude on large scales.  The strong and extended small-scale
clustering reflects more massive parent dark matter haloes and
multiple detected luminous (m$_{i'}<$ 26.4) galaxies having equal and
larger separations per parent dark matter halo on average as compared
to the LBG ACF.  The one-halo term break in the aLBG ACF corresponds
to the virial radii of $\sim$10$^{13} M_\odot$ parent haloes at \zzz\
and is consistent with the higher large-scale clustering amplitude.

In contrast, the eLBG and gLBG ACFs show one-halo term breaks at
$\sim$120 \Mpc, comoving ($\sim$30 kpc, physical), on the same scale
as the full LBG sample, and imply parent halo masses of $\sim$10$^{11}
M_\odot$.  In fact, the gLBG ACF closely follows the full LBG ACF on
all scales.  Although the eLBG ACF traces the LBG ACF reasonably well,
we see an enhancement, or `hump' in the eLBG ACF on intermediate
scales, $\sim$0.5--5 \Mpc.

We note that the LBG subset ACFs show an equivalent, or {\it higher},
amplitude than the full LBG ACF and, thus, the average LBG ACF (the
combination of the three subsets) is less than the sum of its parts.
This result has important implications on the values determined via
correlation function measurements for potentially {\it all} galaxy
populations.  We explore the cause of this effect further via the
spectral type CCFs below and in \S\ref{disc}.

\begin{figure}
\begin{center}
\scalebox{0.38}[0.38]{\rotatebox{90}{\includegraphics{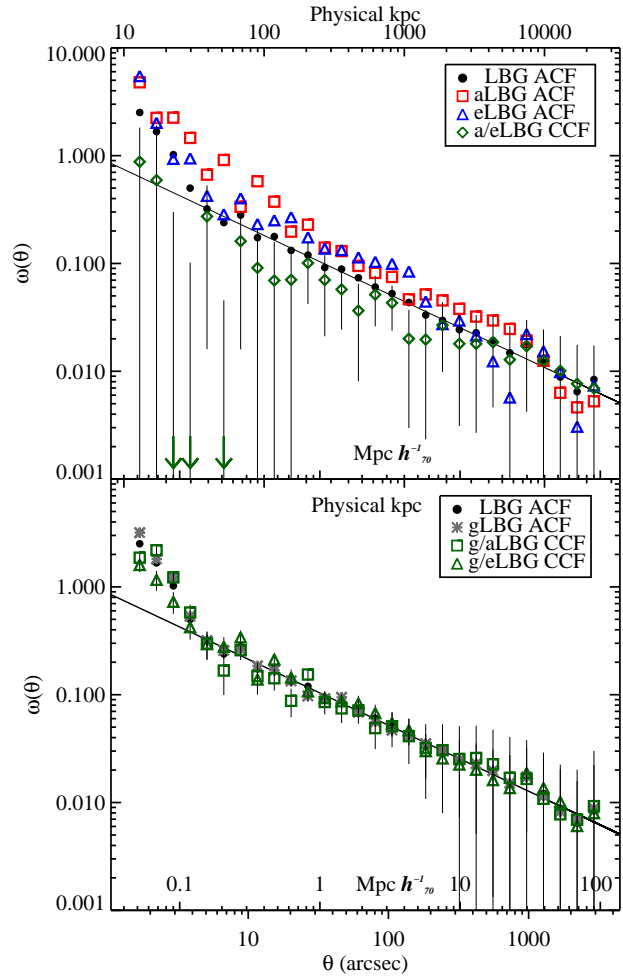}}}

\caption {\small Cross-correlation functions (CCFs) plotted similarly
to Figure~\ref{aelbgs}.  {\bf Upper panel:} The aLBG--eLBG CCF
(diamonds) with the aLBG ACF (squares), eLBG ACF (triangles), and LBG
ACF (solid circles) overlaid without errors for clarity.  The CCF
exhibits an anti-correlation (an amplitude less than the aLBG and eLBG
ACF) indicating that a significant fraction of the two populations
have different physical spatial distributions.  Negative values for
the CCF are indicated with arrows.  {\bf Lower panel:} The gLBG--aLBG
CCF (squares) and gLBG--eLBG CCF (triangles) with the gLBG ACF
(asterisks) and full LBG ACF (filled circles) are overlaid without
errors for clarity.  The gLBG--aLBG and gLBG--eLBG CCF do not show an
appreciable anti-correlation and follow the behaviour of the full LBG
population.}

\label{ccf} 
\end{center} 
\end{figure}

Figure~\ref{ccf} presents the aLBG--eLBG, gLBG--aLBG, and gLBG--eLBG
CCFs.  Bin values of the CCF amplitude that are weaker than the
corresponding ACFs represent an anti-correlation and indicate
different physical spatial distributions.  An anti-correlation occurs
when some fraction of one population does not reside in the same
region of the Universe as the other, such as a location preference for
groups and clusters as opposed to the field and/or as a result of
non-overlapping redshift paths.  The aLBG--eLBG CCF exhibits some
level of anti-correlation at all scales, except the largest separation
bins, and has negative values for three bins within the one-halo
regime (denoted by the arrows in Figure~\ref{ccf}).  In contrast, the
gLBG--aLBG and gLBG--eLBG CCFs show no anti-correlation.  Both CCFs
follow the gLBG ACF and the full LBG ACF within the uncertainties.

The spectral type criteria defined in this work are devised to
generate sub-samples containing a high purity of aLBGs and eLBGs at
the cost of containing all aLBGs and eLBGs.  As a result, the aLBG and
eLBG distributions extend into the gLBG region as is witnessed by our
Keck spectra.  However, for the gLBG--aLBG CCF (gLBG--eLBG CCF) to
show no appreciable anti-correlation implies that the fraction of
eLBGs (aLBGs) in the gLBG region that meet our criteria is relatively
small as compared to the whole and that gLBGs (net \lya\ EW $\sim$ 0)
are found in all environments.

\subsection{Magnitude and Colour Correlation Functions}\label{tests}

One of the main objectives of this work is to examine the behaviour of
LBG sub-sample ACFs based on their spectral type as is motivated by
the observed relationships between \lya\ and other LBG properties.  As
discussed above, the spectral type primary cut makes a diagonal slice
through the CMD that statistically splits the peaks of the aLBG and
eLBG distributions.  Hence, each spectral type includes the effects of
both magnitude and colour.  However, it is equally important, and
highly informative, to examine any effect that magnitude and colour
make on the behaviour of LBG ACFs and to test the effects of different
CMD primary cut slopes.  Here, we divide LBGs into sub-samples in
magnitude and colour to investigate the fundamental drivers behind
various ACF features and, in a coarse sense, the colour and/or
magnitude contribution to the observed differences in the aLBG and
eLBG ACFs.

\subsubsection{Split magnitude and colour correlation
functions}\label{split}

\begin{figure}
\begin{center}
\scalebox{0.38}[0.38]{\rotatebox{90}{\includegraphics{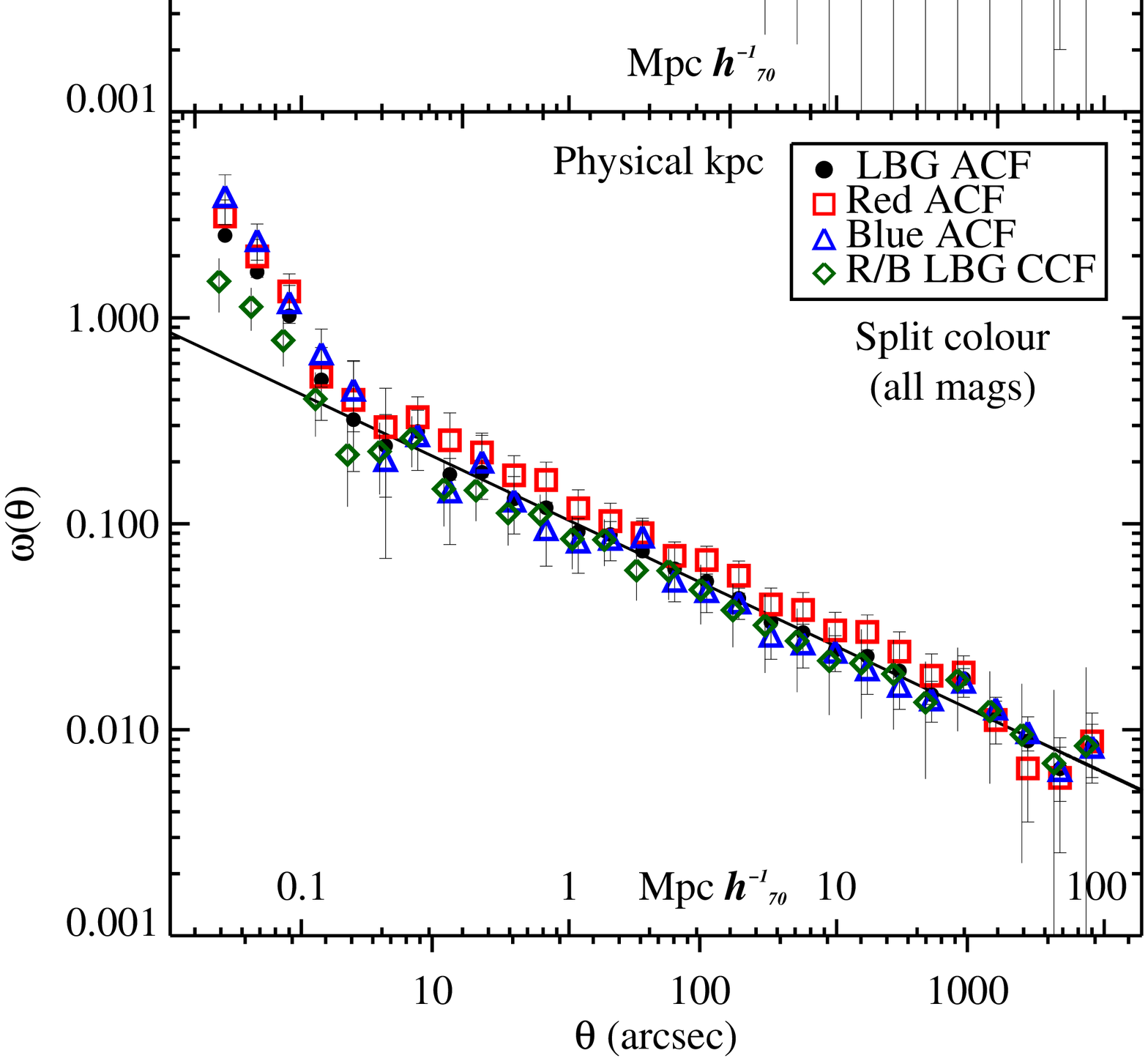}}}

\caption {\small Auto-correlation functions (ACFs) and
cross-correlation functions (CCFs) using split magnitude and split
colour samples that divide the CMD in half {\it (see text)} plotted
similarly to Figures~\ref{aelbgs} \& \ref{ccf}.  The large catalogues
enable accurate correlation functions but contain mixed fractions of
the different spectral types.  {\bf Upper panel:} The split magnitude
`Bright' and `Faint' LBG ACFs.  The `Bright' ACF shows a strong
one-halo term amplitude but roughly agrees with the full LBG ACF over
the two-halo regime whereas the `Faint' ACF behaves more like the eLBG
ACF (see Figure~\ref{aelbgs}) and exhibits a similar enhancement, or
``hump'', around $\sim$0.5--5 \Mpc.  The `Bright--Faint' CCF shows a
small anti-correlation component that is more significant at
intermediate and small scales.  {\bf Lower panel:} The split color
`Red' and `Blue' LBG ACFs.  Both ACFs show similar behaviour and are
nearly identical to the full LBG ACF.  However, the `Red' ACF is
consistently higher than the `Blue' ACF over intermediate to large
scales.  Neither ACF exhibits the eLBG ACF-like ``hump''.  The
`Red--Blue' CCF indicates that the two samples largely co-exist in
space, but shows a anti-correlation component on small scales. }

\label{testplot2}
\end{center}
\end{figure}

As a general examination of the effect that magnitude and colour may
have on the correlation functions, we divide the $(g'-i')$ vs. $i'$
CMD in half at the mean magnitude of the $i'<$ 26.4 LBG sample.  In
this manner, we generate ``split mag'' catalogues containing objects
from the brightest and faintest half of the full LBG sample to
directly test any magnitude effect with a simple non-biased cut.
Similarly, we divide the CMD into ``split colour'' catalogues
containing the reddest and bluest halves of the CMD based on the mean
colour of the full sample.  We compute the ACFs and CCFs for the
``split'' catalogues and present the results in
Figure~\ref{testplot2}.  The sample sizes are large and the
correlation functions can be determined to high accuracy.  However,
each sample contains varying fractions of each spectral type and, in
particular, are dominated in number by gLBGs that may dilute the
contributions from aLBGs and eLBGs.

The mean magnitude of the $i'<$ 26.4 sample is $i'=$ 25.10$\pm0.10$,
1$\sigma$ field-to-field scatter.  When reviewing the CMD, we see that
the split magnitude brighter half, or `Bright' LBG sample, contains
essentially all of the aLBGs, half of the gLBGs (blue and bright), and
the brightest eLBGs.  The `Bright' ACF shows an enhancement in
amplitude over the one-halo term corresponding to haloes of
$M_{DM}\sim$10$^{12} M_\odot$, but weakens to the roughly the
amplitude of the full LBG ACF at larger separations.  The split
magnitude fainter half, or `Faint' LBG sample, contains essentially no
aLBGs, half of the gLBGs (red and faint), and essentially all eLBGs.
The `Faint' ACF follows the full LBG ACF at small scales but then
follows the behaviour of the eLBG ACF at intermediate and large
scales, similarly exhibiting a ``hump'' around $\sim$0.5--5 \Mpc.
Although the `Bright' ACF follows the full LBG ACF at large scales,
interestingly, it is weakest over the range of the ``hump''.

The `Bright--Faint' CCF shows a level of anti-correlation, especially
near the elbow of the one-halo, two-halo terms.  The strongest
anti-correlation coincides with the range of separations in which the
aLBG ACF maintains an enhancement over the `Bright' ACF.  In addition,
the `Bright' and `Faint' LBG samples are quite heterogeneous and a
sufficient fraction of aLBGs and eLBGs, and their extremes, in the two
samples may exist to produce a net anti-correlation.

The mean colour for the $i'<$ 26.4 LBG sample is $(g'-i')=$
0.54$\pm$0.01.  Thus, the split color redder half, or the `Red' LBG
sample, contains the bulk of the aLBGs (the reddest), half of the
gLBGs (red and faint), and essentially no eLBGs.  The split color
bluer half, or the `Blue' LBG sample, contains a small fraction of
aLBGs (the brightest and bluest), half of the gLBGs (blue and bright),
and essentially all of the eLBGs.  The `Blue' LBG sample contains
fewer bright objects as compared to the `Red' LBG sample, as seen in
the natural CMD asymmetry.  The ACFs for both samples closely follow
the full LBG ACF, with the two-halo term amplitude of the `Red' ACF
consistently higher than full LBG ACF and the `Blue' ACF similar to,
or lower than, the full LBG ACF.  Neither ACF appears to show a
``hump''-like feature similar to the eLBG ACF and the `Faint' ACF.
The `Red--Blue' CCF is nearly identical to the full LBG ACF with an
anti-correlation component that becomes significant in the one-halo
term regime.

Interestingly, the split mag samples probe similar redshift paths
(`Bright'; $\langle z\rangle = 3.02, 1\sigma = 0.25$, `Faint';
$\langle z\rangle = 2.93, 1\sigma = 0.35$) as determined by the Keck
spectra, yet show some large scale anti-correlation in the CCF.  The
redshift paths of the split colour samples differ much more (`Red';
$\langle z\rangle = 3.23, 1\sigma = 0.28$, `Blue'; $\langle z\rangle =
2.92, 1\sigma = 0.29$), yet the anti-correlation in the two-halo
regime is small.  The CCFs suggest that the actual redshift paths
probed by the samples are similar enough to only weakly affect the
cross-correlation.  The slope of our spectral-type cut is relatively
flat and bisects the CMD near the mean colour, thus the apparent
small, or lack of, redshift path difference contribution to the
anti-correlation in the split colour CCF is similarly expected for the
aLBG--eLBG CCF.

None of the split samples produce the high amplitude and extent of the
aLBG ACF and the strength of the aLBG--eLBG CCF anti-correlation.
This result shows that the regions defined by net \lya\ EW trace the
LBGs that are generating the extremes.

\subsubsection{Equal magnitude and colour correlation
functions}\label{equal} 

As a complementary test to the split LBG samples and to help assess
the colour and magnitude contributions to the aLBG and eLBG ACFs, we
generate samples with equal magnitudes and colours.  This test carries
the caveat that the data are coarsely binned as a result of the small
samples.

We randomly pull equal distributions of aLBGs and eLBGs from the small
regions where these two spectral types overlap in $i'$ magnitude to
construct `equal mag' samples and in $(g'-i')$ colour to construct
`equal colour' samples.  The `equal mag' samples contain some of the
reddest aLBGs and some of the bluest eLBGs and, as such, we note that
the {\it `equal mag'} samples provide a test of the effects of {\it
colour} on the ACFs.  The distributions are centered at $i'=$ 25.0,
$(g'-i')=$ 1.1 for the `equal mag' aLBGs and $i'=$ 25.0, $(g'-i')=$
0.0 for the `equal mag' eLBGs.  Although the samples have equal
magnitude distributions, they pull from the faintest objects in the
aLBG region and the brightest in the eLBG region.

\begin{figure} \begin{center}
\scalebox{0.38}[0.38]{\rotatebox{90}{\includegraphics{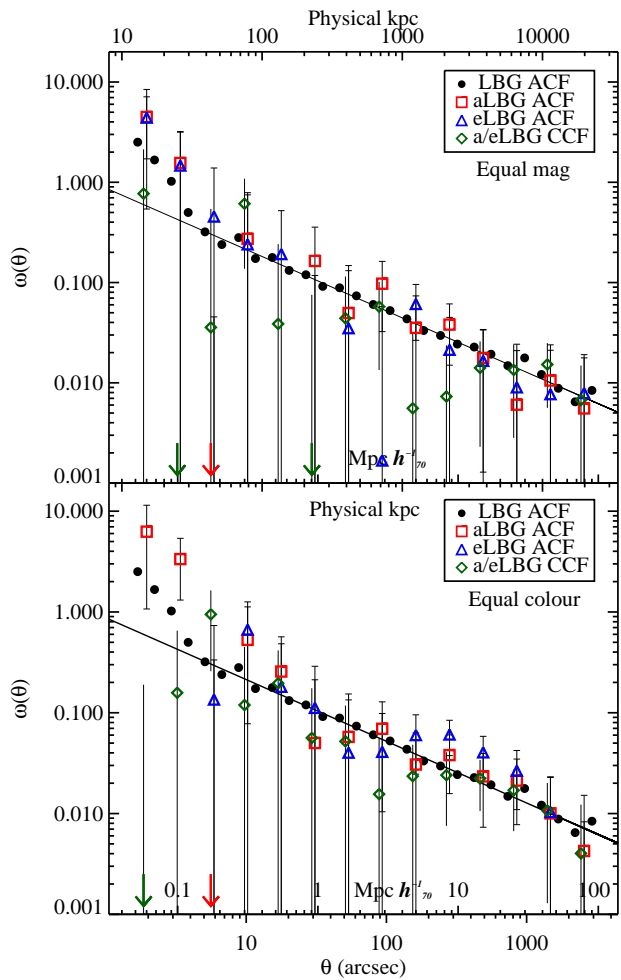}}}

\caption {\small Similar to Figure~\ref{testplot2} but for tests using
spectral type equal magnitude (upper panel) and equal colour (lower
panel) catalogues {\it (see text)}.  The data are plotted with coarser
binning as a result of the relatively small sample sizes.  Values for
the CCFs that are negative are indicated with arrows.  {\bf Upper
panel:} The `equal mag' aLBG and eLBG ACFs and CCF.  Both ACFs appear
to behave similarly and roughly follow the full LBG ACF.  The CCF
exhibits an anti-correlation that suggests that the two populations
may not reside in the same physical locations.  {\bf Lower panel:} The
`equal colour' aLBG and eLBG ACFs and CCF.  The `equal colour' aLBG
ACF shows a strong one-halo term amplitude and, in contrast, the
`equal colour' eLBG ACF shows a negative amplitude.  Both ACFs roughly
follow the full LBG ACF with potentially a higher amplitude on large
scales similar to the `Bright' ACF.  The CCF is negative (strong
anti-correlation) on small scales, helping to support the likelihood
that bright aLBGs and faint eLBGs do not co-exist in the same haloes.
}

\label{testplot1}
\end{center}
\end{figure}

The `equal colour' distributions are centered at $i'=$ 23.0,
$(g'-i')=$ 0.5 for the equal colour aLBGs and $i'=$ 26.1, $(g'-i')=$
0.5 for the `equal colour' eLBGs.  The samples contain some of the
brightest aLBGs and some of the faintest eLBGs and, as such, the {\it
equal colour} samples provide a test of the effects of {\it magnitude}
on the ACFs.  The samples have the same colour distribution but pull
from the bluest aLBGs and the reddest eLBGs.  The ACFs and CCFs for
these samples are presented in Figure~\ref{testplot1}

We find that `equal mag' aLBGs and eLBGs with the same magnitudes have
broadly similar behaviour.  Both ACFs follow the full LBG ACF within
the uncertainties.  Overall, it appears that these two ACFs also
roughly follow the behaviour of the corresponding `Red' and `Blue'
samples in which they are pulled.  There is evidence throughout the
CCF of an anti-correlation suggesting that the two populations, having
the same magnitude but different colour, may not reside in similar
places in the Universe.

The `equal colour' aLBG ACF shows a very strong small scale, one-halo
term, amplitude but, because of the large uncertainties, it is unclear
the extent of the enhancement.  The amplitude appears to weaken with
larger separation and closer to the behaviour of the `Bright' ACF as
compared to the full aLBG ACF.  The `equal colour' aLBG ACF two-halo
term values appear to roughly follow the form of the `Bright' ACF as
well.  The `equal colour' eLBG ACF reveals no enhancement, and instead
a decrement, of galaxies with small separations.  In the two-halo
regime, the ACF roughly follows the `equal color' aLBG ACF.  The lack
of strong anti-correlation in the CCF, except at the smallest scales,
suggests that, if real, the faint eLBGs that make up much of this
sample are not found in the parent haloes of the bright aLBGs but may
exist on the outskirts of the same overdense regions.

We do not have a sufficient number of spectra to determine the
differences in redshift paths probed by the `equal mag' and `equal
colour' samples.  As mentioned earlier, LBGs with higher redshifts are
redder on the CMD.  However, this effect is complicated because of the
inherent differences in colour between aLBGs and eLBGs and because the
two samples have large scatter.  Thus in general, the `equal mag' test
examines the behaviour of small targeted LBG samples with potentially
different mean redshifts whereas the `equal colour' test examines LBG
samples with a potentially similar mean redshifts.

\subsection{Effect of Interlopers}\label{interlopers}

A final consideration is that the clustering of low redshift
interlopers is affecting the form of the ACFs and, in particular, is
driving the strong amplitude of the aLBG ACF.  Some cool Galactic
stars and low-redshift galaxies can meet the \zzz\ LBG color selection
criteria. Our conservative color selection criteria is designed to
minimize the level of contamination.  Here, we review the observed
fractions of low redshift objects and estimate their effects.

We find no Galactic stars in our 68 Keck spectra and S03 using, to a
large extent, similar criteria find $\sim$4\% in their 995 spectra.
Our lower fraction may be due, in part, to our choice of conservative
color-selection criteria in this work which was designed further from
the stellar locus.  The survey of S03 probes to $\cal{R}\lesssim$ 25.5
and our sample extends to $i'\lesssim$ 26.4.  S03 find that the
fraction of stellar contaminants drops to near zero by $\cal{R}\sim$
24 and, thus, would provide little additional contamination in deeper
surveys.  This result may be affected by the difficulty to identify
weak stellar features in faint spectra, but the most distant Galactic
K dwarfs (the faintest main interloper spectral type) estimated in the
directions of the survey pointings are brighter than $\cal{R}\sim$ 24.
As a result, we expect zero to a few percent contamination from
Galactic stars in our $i'<$ 26.4 sample and no coherent clustering
signal contribution.

In certain cases, the 4000\AA\ break and continuum profile of $z\sim$
0.3 galaxies can mimic the drop in flux in \zzz\ LBG continua blueward
of 1216\AA\ from absorption by the \lya\ forest.  To satisfy the
remaining LBG selection criteria, i.e., the drop in flux blueward of
912\AA, the low-redshift galaxies need to be either (1) highly
reddened early type galaxies, (2) star forming galaxies with an
enhancement to their redder broadband colors from strong emission
lines, and/or (3) very faint galaxies that weaken the dynamic range of
the $u^*$ and $g'$ stacked images.  The contamination to our sample
from low-redshift galaxies is estimated to be $\sim$3\% from our Keck
spectra.  This is comparable to the contamination fraction ($\sim$1\%)
of the brighter S03 sample.  From galaxy templates, we find that the
interlopers should populate much of the ($g'-i'$) vs. $i'$ CMD and, in
particular, the central, or gLBG, region, and as such do not comprise
a large enough fraction of any sample to make a noticeable effect on
the ACFs.

The two low-redshift interlopers in our survey have $z=$ 0.343, $g'=$
25.23 and $z=$ 0.356, $g'=$ 25.09 that equate to M$_{B}=$ -15.7 and
M$_{B}\sim$ -16.0, respectively.  The magnitudes probed by our
selection criteria ($i'\sim$22--26.5) give a luminosity range of
M$_B\sim$ -14 to -19.5 and a physical scale of $\sim$5 \kpc\
arcsec$^{-1}$ for galaxies with similar redshifts, whereas \zzz\ LBGs
have M$_{UV}\sim$ -18.5 to -23.5 and a physical scale of $\sim$7.7
\kpc\ arcsec$^{-1}$.  The inflection point where we see the aLBG ACF
depart from a power law, corresponds to the virial radii of
$\sim$10$^{12} M_\odot$ haloes (M$_B\sim$ -20) at $z\sim$ 0.3--0.4.

We plot the colors of our two confirmed low-redshift galaxy
interlopers and find that they fall within, and near, the aLBG
selection region.  If we assume that interlopers do not follow the
template results and exist exclusively in the aLBG region, this
fraction would increase to $\sim$15--20\% of the aLBG population.  If
the low-redshift interlopers are also massive or highly clustered,
this could have the potential to make a measurable effect on the
amplitude and/or form of the aLBG ACF.  However, we find that this is
unlikely for the following reasons.

The two-halo term power law fits to M$_B\sim$ -18 to -20 low-redshift
galaxies is $\gamma\sim$ -0.8
\citep[e.g.,][]{norberg02,lefevre05,zehavi05,li07}.  The fit to the
aLBG ACF is $\gamma=$ -0.63 and in agreement with \zzz\ LBG ACFs in
the literature and our full LBG ACF.

The ACFs of galaxies less luminous than M$_B\sim$ -20 at low redshift
are observed to have very small or no inflections near $\sim$0.2--0.25
Mpc (the inflection point in the aLBG ACF at $z\sim$ 0.35) and more
closely follow smooth power laws down to small scales.  Only galaxies
more luminous than M$_B\lesssim$ -20.5 begin to show an inflection
with the form observed for the aLBG ACF.  Galaxies at $z\sim$ 0.3--0.4
with dark matter haloes of $\gtrsim$10$^{12} M_\odot$ that correspond
to the aLBG ACF inflection are brighter than the brightest end of our
selection magnitude range and would not be selected.  The low-redshift
galaxy interlopers within our magnitude range (M$_B\sim$ -14 to -19.5)
could be sub-haloes to these parent haloes but would also be found
generically in the field.

The interlopers have $i'\sim$ 24 and are not the brightest objects in
our samples.  Fainter interlopers (M$_B\sim$ -14 to -16) have weaker
clustering.  This is likely the case for all interlopers as the
brighter objects in our sample have a higher magnitude dynamic range
between the $u^*$ filter and all others and can be more confidently
selected as high-redshift LBGs via their spectral profile, including
the break in flux blueward of the Lyman limit.  In addition, the
spectra of brighter objects have higher S/N that enables confident
identification of any low-redshift objects.  The fraction of
unidentified $i'<$ 24 spectra in our Keck sample is zero.

Including previous ACF and CCF results and discussions, we conclude
that our defined aLBG ACF reflects closely the true behaviour of \zzz\
aLBGs for the following reasons:

\begin{enumerate} 

\item The fraction of Galactic stars is expected to be very low (we
find zero in our spectra) and any stellar contaminants are expected to
have no coherent clustering signal.

\item The low-redshift galaxy interloper fraction is shown to be small
from our spectra ($\sim$3\%) and the spectra of S03 ($\sim$1\%).

\item Low-redshift galaxies with the luminosities that meet our LBG
selection criteria are observed and theorised to cluster with ACFs
following a $\gamma\sim$ -0.8 power law.  The ACFs of high-redshift
LBGs follow a power law with $\gamma\sim$ -0.6, and the aLBG ACF is
measured to be $\gamma=$ -0.63.  In addition, observations of low
redshift M$_B\lesssim$ -20 galaxy ACFs do not exhibit the strength of
the one-halo term inflection, as is seen in the aLBG ACF.

\item We find no interlopers and no unidentified objects with
$i'\lesssim$ 24 in our spectra.  Objects with $i'\lesssim$ 24 have a
higher dynamic range in the filters and provide more confident LBG
selection.  Thus, interlopers to the sample likely have $i'\gtrsim$ 24
(fainter then $M_{B}\sim$ -16 at $z=$ 0.3--0.4) and are thus low-mass
galaxies.

\item We do not see evidence of enhanced clustering in the eLBG or
gLBG ACFs or any of the various test samples which we would see if a
significant fraction of highly-clustered interlopers occur throughout
the CMD as predicted by galaxy templates and density of objects on the
CMD.

\item We do not see evidence for anti-correlations on large scales in
the CCFs and the test sample CCFs which we would see if a significant
fraction of highly-clustered interlopers are selected by our criteria.

\item If the interlopers are assumed to reside exclusively in the aLBG
region, the split mag `Bright' and split color `Red' samples would
also include the interlopers.  However, these two ACFs do not show
evidence of an enhancement and form from such a population but,
instead, it is divided.  We see a one-halo enhancement in the `Bright'
sample, which includes nearly all aLBGs but also bright eLBGs, and a
two-halo enhancement is seen in the `Red' sample with a consistent
slope $\gamma=$ -0.63, which also includes nearly all the aLBGs but
also faint gLBGs.  In addition, we do not see the corresponding
anti-correlation in the CCFs on large scales.

\item If low-redshift interlopers exist exclusively in the aLBG
region, the aLBG ACF would include an anti-correlation of the two
distinct populations (see \S\ref{emodel} \& \ref{revisit}).  The
anti-correlation component would act to weakening the amplitude over
the two-halo term, but an enhancement is seen instead, unless the
interloper fraction is very large ($\gtrsim$30\%) which our spectra
rule out.  In addition, we would see a two-halo anti-correlation in
the gLBG-aLBG CCF, but we see none.

\end{enumerate}


\section{Analysis and modelling}\label{disc}

\begin{figure} \begin{center}
\scalebox{0.38}[0.38]{\rotatebox{90}{\includegraphics{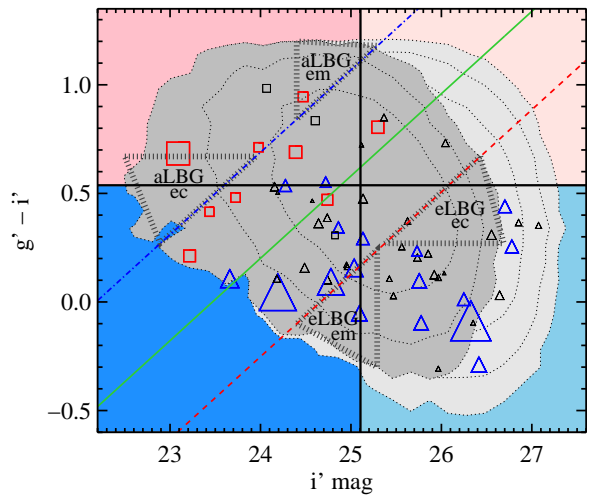}}}

\caption {\small Colour magnitude diagram plotted similarly to
Figure~\ref{264cuts} illustrating the various sections examined.  The
split colour `Red' (top) and `Blue' (bottom) samples and the split
magnitude `Bright' (left) and `Faint' (right) samples are shown
divided by horizontal and vertical solid lines, respectively.  The
labelled triangular thick short-dashed regions denote locations where
the bulk of the `equal mag' aLBGs (top) and eLBGs (bottom) and the
bulk of the `equal colour' aLBGs (left) and eLBGs (right) lay.  The
solid green diagonal line indicates the aLBG--eLBG primary cut.  LBGs
above the blue dot-dashed diagonal line reside $\gtrsim$3$\sigma$ from
the eLBG distribution mean and comprise a nearly pure sample of aLBGs.
LBGs below the red dashed diagonal line reside $\gtrsim$3$\sigma$ from
the aLBG distribution mean and comprise a nearly pure sample of eLBGs.
}

\label{quads}
\end{center}
\end{figure}

The large number of \zzz\ LBGs in the four square-degree CFHTLS Deep
Fields, enable us to break up the CMD into sections to examine the
ACFs for different populations in an effort to better understand the
connection between galaxy UV properties and their spatial
distribution.  Figure~\ref{quads} illustrates the sub-samples in this
work.  We first divided the CMD into three diagonal sections based on
their net \lya\ EW.  We then cut the CMD in half vertically and
horizontally to test the effects of magnitude and colour.  Finally, we
tested small regions that are either common in magnitude or common on
colour to the outer diagonal samples.  The information provided by the
global ACF features enable us to draw several important conclusions
regarding the environment of LBGs with different UV properties, the
haloes in which they reside, and their effect on the measurements of
previous all-inclusive LBG ACFs.

Firstly, we note that the observations here are of the restframe
far-ultraviolet.  Any discussion of ``red'' or ``blue'' LBGs below, or
elsewhere in this work, indicates their placement on the observed
($g'-i'$) versus $i'$ CMD, as all LBGs are starforming or likely have
relatively recent starbursts.  

Secondly, we note that many LBGs with $\lesssim$30 \kpc\ separations
may be interacting.  This includes LBGs in all sub-samples and must be
kept in mind when examining their ACFs and CCFs.  Interaction is known
to induce star formation and strengthen nebular emission line
strengths.  Spectra are necessary to determine whether the close pairs
show \lya\ in emission as is observed for confirmed interacting LBGs
\citep{cooke10}.  Although eLBGs exhibit \lya\ emission by definition,
some interacting LBGs may provide an exception and meet the colour and
magnitude criteria of other spectral types and exhibit \lya\ in
emission as a result of very recent starbursts, given the relatively
short-lived lifetimes of H\textsc{ii} regions and the potential for
escaping \lya\ emission in disturbed systems.

In addition, star formation induced by interactions may boost the
natural magnitudes of faint LBGs and LAEs that would normally fall
just below our detection threshold to above our magnitude limit and
cause them to be included in our samples \citep{berrier12}.  Because
the number density of galaxies increases with magnitude, it may not
take a large fraction of enhanced faint LBGs to produce a measurable
signal in the ACF.  Finally, a fraction of LBGs with small separations
will appear to be close pairs due to projection and the probability of
a projected close pair increases in clustered regions.

The three interacting LBG candidates in our Keck spectra exhibit two
\lya\ peaks, evidence for two closely spaced spectra, and two
corresponding spatially separated sources in the images.  The
candidates are broadly distributed about the centre of the CMD and
reach both the eLBG and aLBG regions.  Thus, any interpretation of the
ACFs of any sub-sample in this work needs to consider that data in the
$\lesssim$30 \kpc\ separation bins likely have some fraction of eLBGs
(net \lya\ EW $\gtrsim$ 20\AA).

\subsection{Examination of the CMD by Quadrant}\label{quadrants}

\subsubsection{`Bright' $\cap$ `Red' quadrant}

The upper left-hand quadrant of the CMD is common to the split mag
`Bright' and split colour `Red' samples.  The two main features of
their ACFs is a strong one-halo term (Bright) and strong two-halo term
(Red).  As such, we see evidence for this corner of the CMD to produce
the highest amplitude ACF on all scales.  This quadrant samples the
bulk of the aLBG region and we see both attributes in the aLBG ACF.
The strength of the one-halo term appears to be dominated by luminous
LBGs whereas the strength of the two-halo term appears to be dominated
by red LBGs.

Combining these results with other ACFs suggests that blue luminous
LBGs have weaker clustering than red luminous, and perhaps red
less-luminous LBGs.  The equal colour aLBG ACF, which focuses on the
most luminous and bluest aLBGs, corroborates this behaviour, although
one must consider the caveats with the small sample sizes and coarse
binning.  Finally, we note that a comparison of the `Red' and `Blue'
ACFs for this purpose needs to consider that the `Red' ACF contains
brighter LBGs on average than the `Blue' ACF because of the natural
asymmetry of the CMD and that each are dominated by fainter LBGs.

\subsubsection{`Faint' $\cap$ `Blue' quadrant}

The split mag `Faint' and split colour `Blue' samples overlap in the
lower right-hand quadrant of the CMD.  Here we find that blue LBGs
have consistently the weakest two-halo term (Blue ACF).  This is the
only ACF to appear weaker on large scales than the full LBG ACF, yet
shows a strong, peaked enhancement at the smallest scales, presumably
due to the brightest members, but is not as strong as that for the
eLBG ACF.  The curious ``hump'' at intermediate scales observed in the
eLBG ACF is also seen in the `Faint' ACF.  In fact, the `Faint' ACF
follows the form of the eLBG ACF over all scales, but is somewhat
diluted and closer toward the form of the full LBG ACF.  The dilution
is expected because the `Faint' ACF includes a significant fraction of
gLBGs whose ACF is nearly identical to the full LBG ACF.

We do not see evidence of a ``hump'' in the `Bright' ACF which
includes bright eLBGs nor the `Red' ACF.  Nor (arguably) do we see any
evidence in either the equal mag eLBG ACF, which includes the
brightest eLBGs or equal colour eLBG ACF, which includes the faintest,
reddest eLBGs.  As a result, we are able to isolate the ``hump''
behaviour to the faintest and bluest eLBGs located in a region of the
CMD that probes LBGs that typically meet LAE criteria.

\subsubsection{`Bright' $\cap$ `Blue' quadrant}

The lower left-hand quadrant is common to the split mag `Bright' and
split colour `Blue' samples.  This region is dominated by gLBGs and
includes approximately equal fractions of the brightest and bluest
aLBGs and eLBGs.  However, because of the asymmetric distribution of
LBGs on the CMD in both colour and magnitude, this quadrant contains
the fewest number of galaxies.  The salient features in both ACFs are
the strong one-halo terms and average to weak two-halo terms.  The
eLBGs in the faint half of the `Blue' sample dominate the ACF and
limit any clear assessment of this quadrant.  Nevertheless, the
observed ACFs, combined with previous quadrant results, further
stresses that luminous LBGs in general have strong one-halo terms
reflecting $\sim$10$^{11-12} M_\odot$ haloes but not necessarily
strong two-halo terms.

\subsubsection{`Faint' $\cap$ `Red' quadrant}

Finally, the split mag `Faint' and split colour `Red' samples share
the upper right-hand quadrant.  Here, we tread in a region of the CMD
where the \lya\ nature of the LBGs is unclear.  The spectroscopic
limits of 8m-class telescopes make identification and EW measures of
\lya\ in absorption of $i'\gtrsim$ 25.5 LBGs extremely difficult and
are not possible with the depths of our Keck spectra.  However,
$i'\gtrsim$ 25.5 LBGs that have \lya\ emission can be identified, and
those with net \lya\ EW $\gtrsim$ 20\AA\ are, by our definition,
classified as eLBGs.  Our spectra find no eLBGs in this quadrant,
three gLBGs with net \lya\ EW $\sim$ 0--10, and one aLBG with net
\lya\ EW $=$ -14.8.  Only the very tip of the eLBG region (faintest,
reddest) and tip of the aLBG region (faintest, reddest) intersect this
quadrant, thus we assume that this area of the CMD contains
predominately gLBGs, an unknown fraction of aLBGs, and little, if any,
eLBGs.

The `Faint' sample is dominated in number by eLBGs and the `Red'
sample by aLBGs and gLBGs.  The limiting magnitudes of the CFHTLS $g'$
and $i'$ images result in the lack of selected objects in the far
upper right-hand corner of the CMD.  Thus the `Faint' and `Red' LBG
ACFs provide little information about the behaviour of LBGs in this
quadrant, however the equal mag aLBG ACF and the equal colour eLBG ACF
probe near, and marginally inside, this quadrant and thus provide a
glimpse of the general behaviour.  Overall, the salient features are
average to weak one-halo term amplitudes and average to strong
amplitudes for their two-halo terms.

\subsubsection{Further CMD Examination}\label{further}

The split mag and split colour samples each contain $\sim$30,000 LBGs.
In addition, the four well-separated square-degree CFHTLS Deep Fields
minimise cosmic variance effects.  Thus the subtleties present in
their ACFs and CCFs may reflect real and distinct features.  Here, we
point out several subtle features that may provide additional
important clues on the spatial distribution of LBGs.

As discussed above, the `Faint' ACF exhibits the same ``hump'' near
$\sim$0.5--5.0 \Mpc\ that appears in the eLBG ACF.  However, the
`Bright' ACF shows a curiously weak amplitude over the same
separations.  Moreover, near $\sim$5 \Mpc, the amplitudes of the
`Bright' and `Faint' ACFs appear to ``switch places''.  The
anti-correlation in the `Bright-Faint' CCF is stronger throughout the
``hump'' region and disappears once the ``hump'' weakens and the
`Bright' ACF increases.  This is in stark contrast to the consistent
amplitudes of the `Red' and `Blue' ACFs and the consistent CCF
correlation over these scales.

The increase in the `Bright--Faint' CCF anti-correlation at
$\sim$0.1--0.5 \Mpc\ separations indicates that these two populations
are generally not found in, or near, each other's parent halo and the
anti-correlation extends to a lesser amount to $\sim$10 \Mpc.  That
is, low-luminosity, and likely low-mass, LBGs are generally not found
near the peaks of overdense regions that host high-luminosity, and
likely massive LBGs, but may reside in the overdense region outskirts.
The anti-correlation may continue to smaller scales ($\lesssim$0.1
\Mpc) but we reach separations in which interactions play a role.

Finally, the consistent anti-correlation strength in the `Red--Blue'
CCF and anti-correlation increase over the one-halo term indicates
that LBGs of each colour typically do not reside in the same places in
the Universe and less so in the same halo.  One interpretation of this
behaviour, given the indications from the above examinations, is that
pairs of red LBGs may occur more often in group-like environments and
pairs of blue LBGs may occur more often near group outskirts or in the
field.

The correlation functions and tests presented in this work illustrate
that only specific sub-samples of LBGs can generate significant
differences in the ACFs and CCFs.  We find that magnitude plays a role
in the strength of the observed one-halo term enhancement but the
extent of the amplitude enhancement is smaller for general samples
than what may be naively expected (cf. the split mag `Bright' sample
ACF).  We also find that colour appears to play a stronger role than
magnitude in tracing more massive haloes, via the two-halo term
amplitude.  Finally, samples probing our defined aLBG and eLBG regions
show the strongest differences in form of all ACFs and the strongest
CCF anti-correlation, potentially lending the greatest insight into
the distribution of LBGs and the environments in which they are found.

\subsection{\lya\ EW and Environment}\label{model}

As discussed earlier, \lya\ EW is a signpost for many LBG properties
including morphology, UV ISM absorption-line strength and velocity
offsets, estimated outflow strength, UV magnitude and colour, and
interaction.  One of the main goals of this work is understand the
spatial distribution of LBGs as a function of net \lya\ EW, to
investigate whether environment plays a role in these observed
relationships.

By definition, the strength of a galaxy ACF at a given separation (ACF
bin) directly describes the prevalence for that galaxy type to exist
at that separation from others of the same galaxy type, after taking
into account any anti-correlation effect.  An obvious example is the
one for typical galaxy ACFs where galaxies are centrally clustered
about specific points in space.  The density of galaxy separations
with respect to random monotonically increases inversely with
separation and the ACF amplitude reveals that information.  Another
example is a galaxy population that clusters in shells about specific
points in space.  Such a geometry would show a more complicated ACF,
as no galaxies are found at the points in space about which the
galaxies cluster (the centres of the shells) and because conventional
ACFs are binned in concentric annuli about each galaxy and most
galaxies would lay near the edges of the shells in projection.
Nevertheless, the geometry can be modelled and discerned from the
shape of the ACF.

The full LBG ACF shows a central clustering behaviour and amplitude
consistent with previous measurements at \zzz\ (Figure~\ref{lbgs}).  A
power law fit to the two-halo term has been shown to correspond to the
clustering of haloes with dark matter masses of
$M_{DM}\sim$10$^{11.5-12} M_\odot$ \citep{adelberger05,cooke06}.  The
inflection point and steeper slope of the one-halo term reflects
average parent haloes having $M_{DM}\sim$10$^{11} M_\odot$ that may
contain more than one luminous galaxy in agreement with previous
findings \citep{ouchi05,lee06}.

\subsubsection{\lya\ EW $\sim$ 0\AA\ (gLBGs)} 

The gLBG selection region forms a thick diagonal band across the
centre of the CMD and thus the bulk of gLBGs sample LBGs of average
colour, magnitude, and \lya\ EW.  The gLBG ACF closely follows the
full LBG ACF over all scales (Figure~\ref{aelbgs}; bottom panel) and
implies that a large fraction of galaxies meeting this spectral type
criteria are found in average LBG haloes.  The lack of an
anti-correlation in the gLBG--aLBG or gLBG--eLBG CCF implies that
gLBGs exist to some extent in all environments discussed below.

\subsubsection{\lya\ EW $\lesssim$ -10\AA\ (aLBGs)} 

The aLBG ACF is also centrally clustered, but displays a consistently
higher amplitude as compared to the full LBG ACF (Figure~\ref{aelbgs};
top panel).  Although the aLBG ACF one-halo term central values have
scatter, they remain higher than the full LBG ACF out to approximately
the virial radii of haloes with $M_{DM}\sim$10$^{13} M_\odot$.  In
addition, the amplitude of the two-halo term is roughly 1.5$\times$
that of the full LBG ACF and is not inconsistent with haloes of this
average mass.  We investigate the aLBG ACF in more detail in a future
paper, however, the observed behaviour of the aLBG ACF, and that of
other LBG sample ACFs and CCFs, lead to the conclusion that massive,
group-like haloes preferentially contain aLBGs.

\subsubsection{\lya\ EW $\gtrsim$ 20\AA\ (eLBGs) and the shell
model}\label{emodel}

The eLBG ACF shows a centrally clustered behaviour but includes the
curious ``hump'' in amplitude over $\sim$0.5--5 \Mpc\ and subsequent
drop from $\sim$5--25 \Mpc\ that we also see in the `Faint' LBG ACF.
As a reminder, the `Faint' LBG sample is dominated, in number, by
eLBGs.  The eLBG ACF one-halo terms does not resemble that of the aLBG
ACF.  Instead, it displays an inflection near $\sim$30 \kpc\
($\sim$0.12 \Mpc), similar to the full LBG ACF, corresponding to
parent dark matter haloes of $M_{DM}\sim$10$^{11} M_\odot$ with a
steep peak to the smallest scales.

Because both the eLBG and `Faint' LBG ACFs exhibit the ``hump''
feature, and because both samples contain a large number of LBGs, the
observed form of these ACFs is very likely real and motivates a
modeling of a geometry that might cause such a spatial distribution.
The results of the various ACF and CCF analyses in this work show that
eLBGs typically do not have a strong one-halo term and the enhancement
in their ACF between $\sim$0.5--5 \Mpc\ suggests that there is an
overabundance of eLBGs on these scales.  Consequently, we investigate
a model with a geometry in which galaxies are placed exclusively at
these scales, termed the ``shell'' model.

\begin{figure} \begin{center}
\scalebox{0.38}[0.38]{\rotatebox{90}{\includegraphics{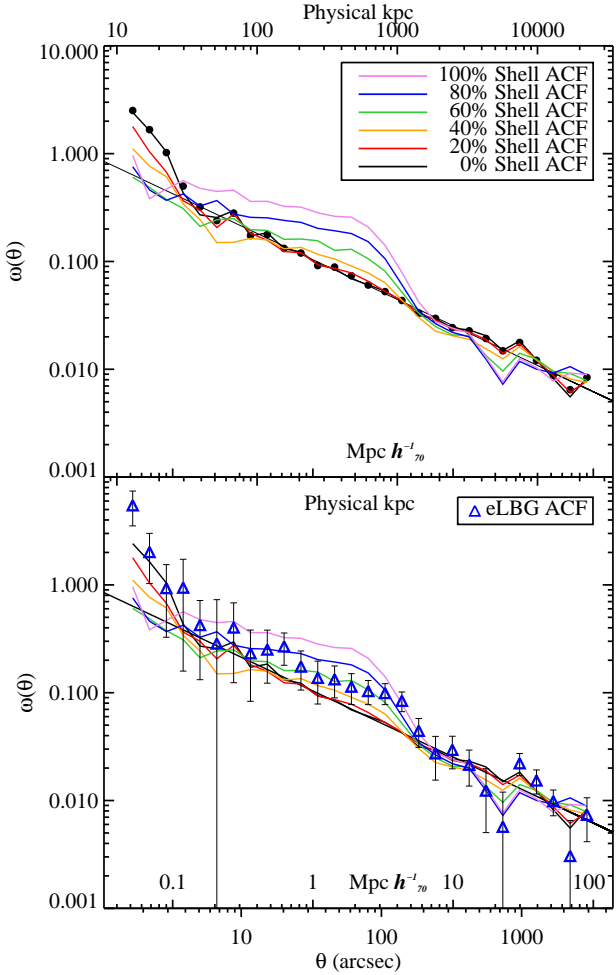}}}

\caption {\small Simulated and observed auto-correlation functions
(ACFs) for Lyman break galaxies (LBGs) with different spatial
distributions.  {\bf Top panel:} The observed centrally-clustered LBG
ACF (solid circles) is shown.  Overlaid are the ACFs for six
populations that consist of LBGs with 0\%, 20\%, 40\%, 60\%, 80\%, and
100\% fraction of simulated galaxies residing exclusively in 2--4
\Mpc\ shells {\it (see text)}.  {\bf Bottom panel:} The observed eLBG
ACF with the ACFs for the six populations overlaid for comparison.  A
population of LBGs that contains $\sim$60\% of its members in 2--4
\Mpc\ shells appears to reproduce the unusual two-halo term behaviour
of the eLBG ACF.}

\label{shell}
\end{center}
\end{figure}

We place galaxies randomly on spherical shells with radii, guided by
the form of the eLBG ACF, ranging from 2--4 \Mpc\ and randomly
distribute the shells of galaxies in the CFHTLS fields.  We place 10
galaxies on each shell and then match the total number of galaxies to
the number of LBGs in each CFHTLS field.  Figure~\ref{shell} shows the
resulting ACF (violet curve; 100\% Shell ACF).  The main feature of
the shell ACF is an increasing amplitude from small to intermediate
scales, with a peak near the shell diameters.  All separations are the
projected separations of the 3-D shells and, for such a geometry, the
density of galaxies increases near the edges of each 2-D projected
shell.  Because the ACF is computed in logarithmic annuli about each
galaxy, the largest number of pair separations exist on scales that
range roughly from the radius to diameter of each shell.

A population residing exclusively on shells will show a decrease in
amplitude on scales larger then the shell diameters (here
$\gtrsim$2--4 \Mpc) and, depending on the density of shells, can
decrease below that of a centrally-clustered population as a result of
the space between shells, as can be seen in Figure~\ref{shell}.  This
leads to a ``dip'' in the ACF and, for our model, occurs on scales of
$\sim$10--20 \Mpc, which is slowly recovered at the largest
separations as pairs become regularly sampled between independent
shells.  We find that the dip is persistent when testing smaller and
larger radii shells.  The shell ACF reproduces the form of the eLBG
ACF two-halo term, naturally producing the intermediate-scale ``hump''
and subsequent ``dip'' in amplitude.  However, the ``hump'' amplitude
is much higher than that seen in the eLBG ACF.  The one-halo amplitude
is not reproduced and would result from either a mixture of shell and
LBG-like centrally clustered distributions or nearly exclusively from
interactions.

We then compute the ACFs for LBG populations containing various
fractions of its galaxies in shells.  We use the full LBG data to
model a ``normal'' centrally clustered population (0\% shell
population) and replace 20\%, 40\%, 60\%, and 80\% of the data with
the simulated shell galaxies.  Figure~\ref{shell} presents the
results.  We see the form of the two independent ACFs slowly merge in
a non-linear manner, however, which is discussed in \S\ref{mass}.  We
find that the ACF of a population with $\sim$50--70\% of its members
with a shell or shell-like distribution is able to describe the
curious eLBG ACF two-halo term well.  The lower panel of
Figure~\ref{shell} compares the LBG ACFs with varying fractions of
shell members to the eLBG ACF.  We note that the shell model was
arbitrary designed to have 2--4 \Mpc\ shells as a test of concept and
the true distribution is likely different.  However, if the eLBG
population is indeed comprised of a fraction of members in shells,
given the form of the mixed population ACF, the average true range of
shell radii is not too different.

\subsubsection{An emerging picture}\label{picture}

The aLBG and eLBG ACFs produce the largest differences of any LBG
sub-sample pair tested here and demonstrate that the two populations
behave very differently.  Furthermore, the strength of the aLBG--eLBG
CCF anti-correlation is not duplicated by any other sub-sample CCF and
indicates that, on average, the two populations reside in decidedly
different environments.  The extent of the aLBG ACF amplitude
enhancement implies that aLBGs largely exist within massive,
$\sim$10$^{13} M_\odot$, group-scale parent haloes.  The eLBG ACF
one-halo term enhancement reflects typical LBG halo masses
($\sim10^{11} M_\odot$) and a two-halo term that shows a potential for
a shell-like geometry for a significant fraction of its members.  The
scale of the two-halo enhancement, reinforced by the comparison to our
shell models, reflects shell sizes corresponding to radii in a range
near $\sim$2--4 \Mpc.

We can also infer that few luminous LBGs reside in this separation
range as evidenced by the reverse ``hump'' behaviour in the `Bright'
ACF (comprised largely of aLBGs and gLBGs) and the consistent
anti-correlation in the `Bright--Faint' CCF from $\sim$1--5 \Mpc.  In
addition, we see the strongest anti-correlation from $\sim$0.1 to
nearly 1 \Mpc\ that may support the lack of faint LBGs in massive
parent haloes.

The faintest, bluest eLBGs were found to be responsible for the
unusual form of the eLBG ACF.  Assuming the ``hump'' in the eLBG ACF
results from galaxies with shell-like distributions, and to remain
consistent with the expectations of $\Lambda$CDM cosmology, we propose
that a significant fraction of faint, blue eLBGs reside on the
outskirts of massive haloes hosting luminous LBGs.  Red aLBGs are
found throughout the most massive haloes and are, thus, likely hosted
by very luminous aLBGs.  The radii of the shells in the model are not
a best fit to the data, but instead are only values guided by the
scaling of the ``hump'' in the eLBG ACF.  The $\sim$2--4 \Mpc\ radii
used in the shell model equate to, or are larger than, the most
massive haloes that exist at \zzz.  Thus, many of the ``shell'' eLBGs
are likely on the outskirts and outside massive haloes, perhaps in
connecting filamentary structure and/or possibly infalling.  This
picture is consistent with all ACFs and CCFs in this work, including
the anti-correlation in the aLBG--eLBG CCF.

Along with dominant \lya\ emission and blue continua, typical eLBGs
have weak/narrow ISM absorption features and compact morphology
\citep{shapley03,law07,cooke10,law11}.  Fainter LBGs have lower galaxy
bias suggesting that the eLBGs dominating the shells are low mass
galaxies.  In addition, gamma ray burst studies indicate that hosts
meeting eLBG criteria have lower than average metallicity
\citep[e.g.,][]{chen09}.  Perhaps group outskirts provide conditions
for low mass haloes to undergo efficient or induced star formation,
possibly their first star burst, that makes the galaxies readily
detectable with the LBG colour selection technique.  At closer
proximity to the centers of massive haloes, low mass galaxies may
experience effects from the denser environment, such as ram pressure
stripping and harassment, resulting in lower gas fraction/star
formation efficiency and either elude LBG colour selection detection
or evolve and become gLBGs or aLBGs.  Typically, aLBGs are larger in
extent, more diffuse, and often contain multiple clumpy small star
forming regions.  Because we find aLBGs preferentially in massive and
group-like haloes, these may be more mature galaxies and/or the
results of earlier mergers.

The overall behaviour of aLBGs and eLBGs is reminiscent of the local
morphology-density relation \citep[e.g.,][]{dressler80}.  If this
picture is confirmed, the properties of aLBGs and eLBGs, and in
particular \lya, provide a spectroscopic and morphological means to
trace environment and help understand the assembly of groups and
clusters in the early Universe.

\subsection{Underestimated mass estimates}\label{mass}

The aLBG--eLBG CCF anti-correlation helps explain how the aLBG and
eLBG ACF amplitudes are higher than the full LBG ACF and reinforces
the distinct nature of the two LBG spectral types.  But the ACFs and
CCF amplitudes have an additional important implication.  By
definition, the full LBG ACF is comprised of aLBGs, eLBGs, and gLBGs.
However, the amplitudes of the ACFs for the three sub-samples are
equivalent to or higher than the full LBG ACF, with none being lower
(cf. Figure~\ref{aelbgs}).

This result empirically shows that the ACF amplitude for the LBG
population taken as a whole, regardless of spectral type (which has
only been done to date), is lower than the true average, indicating
that the correlation length and average mass of LBGs has been
underestimated.  The effect is driven by the anti-correlation of the
sub-samples and applies to other galaxy populations if they, too, are
shown to have distinct populations that exhibit spatial segregation or
differing clustering behaviour.

\begin{figure} \begin{center}
\scalebox{0.46}[0.44]{\includegraphics{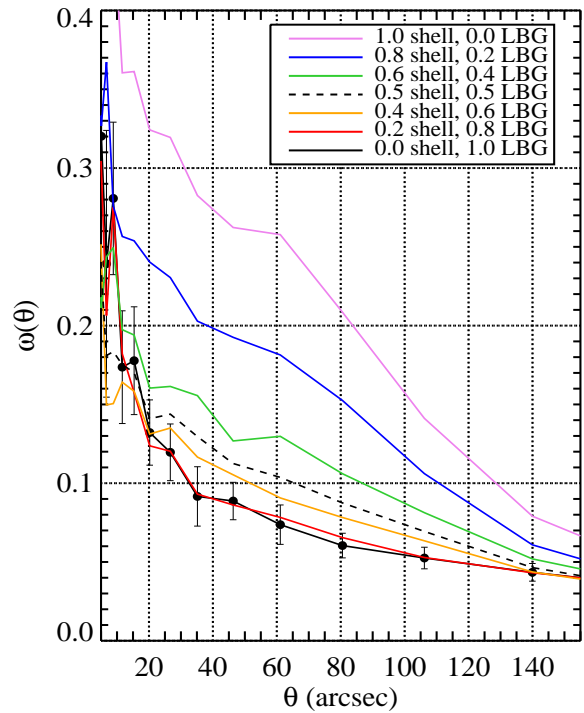}}

\caption {\small Zoom-in of Figure~\ref{shell}, upper panel, plotted
in linear space.  The auto-correlation functions of the mixed shell
and LBG populations show a non-linearity in the amplitude as a
function of mixing fraction.  The dashed curve has been added to show
the auto-correlation function for a population that includes 50\% of
each type.  The amplitude of the 50/50 population is not he average of
the two independent auto-correlation functions.  Because the two
galaxy types have different spatial distributions, their
anti-correlation weakens the averaged auto-correlation amplitude.
Uncorrected, this effect would produce lower average masses for the
full population. }

\label{linear}
\end{center}
\end{figure}

The analysis in \S\ref{emodel} is a test of this effect.  The ACFs are
comprised of two populations with very different spatial
distributions.  The ACF amplitudes of the mixed samples (shell and
centrally clustered distributions) do not reflect the expectations of
the averages of the independent ACFs.  This point is illustrated in
Figure~\ref{linear}.  For example, the shell sample ACF amplitude is
seen to drop more than 20\% relative to the LBG ACF when introducing a
20\% fraction of centrally-clustered LBGs (that is, when comparing the
difference between the 1.0 and 0.0 Shell ACFs to the 0.8 and 0.0 Shell
ACFs).  In fact, the drop is 42.3$\pm$3.9\% when averaged between
20--140 arcsec.  The drop in amplitudes when comparing the three other
Shell ACFs to the 1.0 Shell ACF are 73.6$\pm$6.6\% (0.6 Shell ACF),
91.6$\pm$4.8\% (0.4 Shell ACF), and 99.5$\pm$2.7\% (0.2 Shell ACF).

The larger drop in amplitude with respect to the expected dilution
from averaging the two ACF amplitude values can be attributed to the
anti-correlation component of the two spatially distinct populations
being picked up by the joint ACF.  Because we would naively expect 20,
40, 60, and 80\% drops in amplitude, but find 42.3, 73.6, 91.6, and
99.5\%, the anti-correlation component provides an estimated
27$\pm$7\% negative contribution to the average ACF amplitude.

The above test is based on our simple shell ACF model in comparison to
the observed full LBG ACF.  Our data show that the aLBG ACF follows
the centrally clustered form and has a 41.1$\pm$15.6\% higher
amplitude than the full LBG ACF over the two-halo term separation
range $\sim$1--10 \Mpc.  The eLBG ACF has a 49.8$\pm$31.7\% higher
amplitude over $\sim$1--10 \Mpc\ and a 68.9$\pm$22.4\% higher amplitude
over 20--140 arcsec where the presumed shell geometry is the
strongest.  Because the gLBG ACF is consistent with the full LBG ACF
and the gLBG--aLBG and gLBG--eLBG CCFs show little anti-correlation,
the anti-correlation needed to reduce the aLBG and eLBG ACF amplitudes
to that of the full LBG ACF is $\sim$45\%.  Examining the aLBG--eLBG
CCF, we find that the CCF reflects a 29.7$\pm$15.5\% drop from the
full LBG ACF amplitude over $\sim$1--10 \Mpc.

Although our conservative spectral type criteria identify 21,215 aLBGs
and eLBGs (37\% of the full sample), approximately 50\% of the LBG
population meet the net \lya\ EW cuts.  The additional 10--15\% fall
in our defined gLBG region.  Depending on whether or not these gLBGs
behave similarly to aLBGs and eLBGs, and realising that the aLBGs and
eLBGs may have a small level of anti-correlation across each sample,
the full contribution could range from $\sim$15 to perhaps $\sim$40\%.
A 15--40\% increase in the ``true'' amplitude of the full LBG ACF
would roughly correspond to an increase in the correlation length,
r$_0$, of the \zzz\ LBG population from r$_0\sim$4.0
\citep{adelberger05} to r$_0\sim$4.4--5.0 and an inferred mass 1.5--3
times greater than currently estimated.

Clearly, this effect is not limited to LBGs and is inherent to the
correlation function formalism which assumes a homogeneous population.
As a result, the anti-correlation contribution for any galaxy
population having sub-samples that have different spatial distribution
may be significant and needs to be considered when computing and
inferring values from the average correlation functions of entire
galaxy populations.

\subsection{Interlopers revisited}\label{revisit}

We return to the issue of the effect low-redshift interlopers may have
on our LBG ACFs and, in particular, the high amplitude of the aLBG
ACF.  In section \S\ref{interlopers}, we detail the reasons why our
sample cannot have a significant interloper fraction.  In addition to
those reasons, we mention that the low redshift population would be
spatially distinct from the high redshift population and would generate
an anti-correlation component in the aLBG ACF such that it would
require an interloper fraction well beyond that allowed by our spectra
to make a significant affect on the amplitude.

Reviewing the observed and simulated test ACFs, we see that it
requires $\gtrsim$30\% contamination of the shell population to
generate any measurable increase in amplitude from the observed LBG
ACF to counter the effect of the anti-correlation component.  Both the
test ACFs and the aLBG, gLBG, and eLBG ACFs demonstrate that an ACF
for a full population includes the anti-correlation component inherent
to that population from members that have different spatial
distributions.

If the magnitude of the effect of the shell ACF on the full LBG ACF is
similar to that of a strongly clustered low redshift population, it
would take a similar, if not larger, fraction of low-redshift
interlopers to generate the high amplitude observed for the aLBG ACF.
Our Keck spectra rule out any fraction greater than $\sim$5\%, and
$\sim$20\% when we make the extreme assumption that all interlopers
are localised exclusively in the aLBG region.  As can be seen in
Figure~\ref{shell}, the effect of a contamination of $\sim$20\% is
negligible on the amplitude of the ACF as a result of the inclusion of
the spatial anti-correlation.


\section{SUMMARY}\label{conc}

We identify $\sim$57,000 \zzz\ LBGs to a limiting magnitude of
$i'\sim$ 26.4 in four square-degree CFHTLS Deep Field stacked images.
Our conservative colour-selection criteria follows that of other
successful surveys and are demonstrated to select a clean and
representative LBG population, confirmed by our 68 Keck spectra.  The
large sample size and square-degree fields enable the measurement of
an accurate LBG auto-correlation function (ACF) from small to large
scales and accurate ACFs and cross-correlation functions (CCFs) for
various sub-samples explored here.

Motivated by the diagonal gradient of net \lya\ EW across the CMD and
the relationships of \lya\ EW with other UV spectral features, colour,
interaction, and morphology, we divide the CMD into sections to select
$>$95\% pure samples of LBGs having dominant net \lya\ EW in
absorption (aLBGs), dominant net \lya\ EW in emission (eLBGs), and net
\lya\ EW near zero (gLBGs).  In addition, we divide the CMD in half
vertically and horizontally to explore the effects of colour and
magnitude on the ACFs and CCFs.  We summarise our results as follows.

\begin{itemize}

\item We find the two-halo term of the full LBG ACF closely follows a
power law and is consistent in amplitude and slope with previous work.
We fit a power law of the form $\omega (\theta) = A~\theta^\gamma$ to
the data from $\sim$1--20 \Mpc\ and find $\gamma=$ -0.61.  We find a
departure from a power law at $\sim$0.12 \Mpc\ ($\sim$30 \kpc,
physical), similar to that found at $z\sim$ 4 by \citet{ouchi05}
probing LBGs with similar luminosities over similar scales.  The break
in the ACF corresponds to the virial radii of haloes of
$M_{DM}\sim10^{11} M_\odot$.  The steep rise in amplitude of the
one-halo term on the smallest scales suggests that LBG haloes contain
multiple luminous, $i'\lesssim$ 26.4, galaxy sub-haloes and/or
reflects interacting pairs.

\item The aLBG ACF exhibits a strong one-halo term amplitude extending
to $\sim$0.8--1 \Mpc\ ($\sim$200--250 \kpc, physical) and
corresponding to the virial radii of $M_{DM}\sim10^{13} M_\odot$
haloes.  The amplitude of the two-halo term is consistently higher
than that of the full LBG ACF and in agreement with expectations for
more massive haloes.

\item The eLBG ACF shows a one-halo term inflection point consistent
with that of the full LBG ACF and, thus, similarly implies typical
haloes of $M_{DM}\sim10^{11} M_\odot$.  The eLBG ACF two-halo term
shows a curious ``hump'' from $\sim$0.5--5 \Mpc\ where it is
significantly higher than the full LBG ACF and then exhibits a drop
between $\sim$5--25\Mpc.

\item The gLBG sample contains the largest number of members and
includes the bulk of LBGs with average colour and magnitude.  We find
that the gLBG ACF is nearly identical to the full LBG ACF.

\item The aLBG--eLBG CCF show a strong anti-correlation component over
all scales, except the largest, suggesting that a significant fraction
of the two populations do not reside in the same physical
locations/environments.  In contrast, the gLBG--aLBG and gLBG--eLBG
CCFs show no appreciable anti-correlation component.

\item Splitting the CMD in half in magnitude, we find that the ACF for
`Bright' LBGs has a strong one-halo term, corresponding to
$M_{DM}\sim10^{11.5-12} M_\odot$ haloes, but a two-halo term that is
only marginally stronger than the full LBG ACF.  The `Faint' LBGs ACF
has a one-halo term similar to the full LBG ACF and exhibits the
``hump'' feature seen the eLBG ACF.  Finally, the CCF for the two
samples shows a weak anti-correlation over $\sim$1--20 \Mpc,
increasing in strength near the inflection point between the one- and
two-halo terms.

\item Splitting the CMD in half in colour, we find that the ACFs for
both the red and blue LBGs have roughly average one-halo term
amplitudes but the red LBG ACF has a strong two-halo term that
consistently remains $\sim$1.5$\times$ stronger than the full LBG ACF
from $\sim$0.2--20 \Mpc, whereas the blue LBG ACF is consistently weak
over the same scales.  We see little anti-correlation in the CCF,
except at the smallest scales $<$0.1 \Mpc\ ($<$25 \kpc).

\item The eLBG sample consists of $\sim$12,000 galaxies and the
`Faint' LBG sample consists of $\sim$29,000 galaxies.  We see the
unusual ``hump'' feature in both the eLBG ACF and `Faint' LBG ACF,
thus, the feature is likely real.  The feature is not seen in the
other adjacent ACFs and is localised to the faintest, bluest LBGs.
Based on the results from all ACFs and CCFs in this work and our
examination of the CMD by quadrant, we test a model of eLBGs that
includes a significant fraction of galaxies residing exclusively on
shells.  We find that such a model reproduces the ``hump'' from
$\sim$0.5--5 \Mpc and the decrease in amplitude from $\sim$5--25 \Mpc\
seen in the eLBG and `Faint' LBG ACF.  If the real eLBG distribution
contains galaxies having a similar geometry, then we find that
$\sim$60\% of eLBGs are in shell-like structures with roughly
$\sim$2--4 \Mpc\ radii.

\item Finally, we find that the aLBG, eLBG, and gLBG sub-samples have
equivalent or higher ACF amplitudes than the full LBG sample ACF in
which they are pulled.  In other words, the amplitude of the full LBG
ACF is weaker than the sum of its parts.  The anti-correlation
component in the aLBG--eLBG CCF as a result of their differing spatial
distributions acts to weaken the ACF amplitude when averaging the full
population.  Based on our simulated galaxies and the data, we estimate
the anti-correlation component decreases the two-halo term ACF
amplitude by $\sim$15--40\% indicating that the ``true'' inferred mass
of \zzz\ LBGs is $\sim$1.5--3 times greater than previously measured.
The results suggest that ACFs determined for any galaxy population
that consists of members with different spatial distributions, i.e.,
members that reside in different environments, will always
underestimate the true average amplitude of the population.  The
effect can be significant and needs to be considered in future work.

The correlation functions and tests presented in this work illustrate
that only specific sub-samples of LBGs can generate significant ACF
and CCF differences.  We find that magnitude plays a role in the
strength of the observed one-halo term enhancement and that colour
appears to play a stronger role than magnitude in the two-halo regime.
We find that samples probing our defined aLBG and eLBG regions show
the largest ACF differences and the strongest CCF anti-correlation,
potentially lending the greatest insight into the distribution of LBGs
and the environments in which they are found.

The LBG spectral type results in this work, based on net \lya\ EW
(aLBG, eLBG, and gLBG ACFs and CCFs), are corroborated by the ACFs and
CCFs of the unbiased colour and magnitude samples and smaller test
samples that are comprised of aLBGs and eLBGs with equal colour and
equal magnitude distributions.  Taken in whole, the results point to a
picture where aLBGs are preferentially located in massive, group-like
environments and eLBGs are located on halo and group halo outskirts
and in the field.  Because net \lya\ EW which is known to trace many
other intrinsic properties, including star formation rate and
morphology, the behaviour of the spectral types presented in this work
demonstrate that the mechanisms behind the morphology-density relation
at low redshift are in place at \zzz\ and implies that LBG UV
spectroscopic features, in particular, \lya, may be a strong indicator
of environment.

\end{itemize}


\section*{Acknowledgments}

The authors would like to thank M. Ouchi and J. Koda for helpful
discussions.  ERW acknowledges the support of Australian Research
Council grant DP 1095600.  The results presented here are based on
observations obtained with MegaPrime/MegaCam, a joint project of CFHT
and CEA/DAPNIA, at the Canada-France-Hawaii Telescope (CFHT) which is
operated by the National Research Council (NRC) of Canada, the
Institut National des Science de l'Univers of the Centre National de
la Recherche Scientifique (CNRS) of France, and the University of
Hawaii. This work is based in part on data products produced at
TERAPIX and the Canadian Astronomy Data Centre as part of the
Canada-France-Hawaii Telescope Legacy Survey, a collaborative project
of NRC and CNRS.  In addition, the results rely on observations
obtained at the W. M. Keck Observatory.  Both the CFHT and W. M. Keck
Observatory are located on the summit of Mauna Kea, Hawaii.  The
authors wish to recognise and acknowledge the very significant
cultural role and reverence that the summit of Mauna Kea has always
had within the indigenous Hawaiian community.  We are most fortunate
to have the opportunity to conduct observations from this mountain.


\label{lastpage}

\end{document}